\documentclass[12pt]{article}

\usepackage{latexsym}
\usepackage{graphicx}

\begin{document}
\begin{center}{\large Noise-induced Input Dependence in a
Convective Unstable Dynamical System}
\end{center}

\begin{center}
Koichi FUJIMOTO and Kunihiko KANEKO

{\footnotesize {\it Department of Pure and Applied Science, University of Tokyo, 
Komaba, Meguro-ku, Tokyo, 153-0041, Japan}}
\end{center}

\begin{abstract}
Unidirectionally coupled dynamical system is studied by focusing on the input (or 
boundary) dependence. Due to convective instability, noise at an up-flow is spatially 
amplified to form an oscillation. The response, given by the down-flow dynamics, shows 
both analogue and digital changes, where the former is represented by oscillation 
frequency and the latter by different type of dynamics. The underlying universal 
mechanism for these changes is clarified by the spatial change of the co-moving 
Lyapunov exponent, with which the condition for the input dependence is formulated. 
The mechanism has a remarkable dependence on the noise strength, and works only 
within its medium range. Relevance of our mechanism to intra-cellular signal dynamics 
is discussed, by making our dynamics correspond to the auto-catalytic biochemical reaction 
for the chemical concentration, and the input to the external signal, and the noise 
to the concentration fluctuation of chemicals. 

\end{abstract}

\section{Introduction}
In a biological system, signaling phenomena are important, that 
transform external inputs to outputs. In general the signaling process is not simple, and 
is not given by a fixed one-to-one correspondence. A typical example is an intra-cellular 
signaling, where several chemicals are involved. These chemicals include several 
catalytic reactions and positive feedback process to amplify the input. As a result 
nonlinear dynamics of coupled elements is concerned. 

In a cell, an external stimulus coming from the outside of a cell is selected, amplified, 
and transported as a signal. Recent development in molecular biology provides us more 
and more detailed information on the signaling process\cite{theCell}. In the experiment 
of molecular biology, mechanisms on chemical signal transmission have become clearer. 
Network of the pathways from a given stimulus through signal chemicals to the final 
response is also being clarified. 

The network of signaling pathway is generally very much complicated. A large number 
of biochemical reactions are involved, from the cell membrane to ligand, and to the final 
response reaction. Often, the pathway may not be uni-directional, but several pathways 
interfere with each other. With this complication one could even expect that there would 
be no correlation between input and final response. If such were the case, however, no 
biological function would be expected. Then, why is the network of signal pathway so 
long and complicated? How can the signal pathway respond suitably to inputs? How 
does the signal process work within possible thermodynamic fluctuations? 

To answer such general questions, experimental molecular biology is not sufficient, but 
some theoretical study on general features of a biosystem is required. To address the 
above questions, we start from a study of simple model pathway formed uni-directionally, 
as an ideal simplification of the signal pathway, where a model with a sequence of 
biochemical reactions from an input is adopted, that transforms input to the final 
response. Here we do not take a biologically realistic model, but focus on a dynamical 
system aspect abstracted by a simple model. With this we will propose a mechanism 
how a signal is amplified and transmitted within fluctuations. 

To understand the mechanism of signaling, we note that the signal system should 
satisfy at least the following three properties;

\begin{description}
\item[property 1] : Amplification and transmission of input ; external signal is 
amplified and transmitted into an intra-cellular dynamics. 
\item[property 2] : (Linear) stability : the system returns to the original state when the 
input is off. 
\item[property 3] : Input dependence : response following the internal dynamics  
depends on the nature and strength of the input signal. 
\end{description}

In the present paper, we introduce an abstract model with a simple biochemical signal 
pathway, and show that its dynamics can satisfy the above three properties. A chain of 
coupled biochemical reactions represents the model. Indeed, in signal pathways, there 
are several chains of reactions, where by means of enzyme-catalyzed pathways, many 
biochemical reactions are driven in one direction, through coupling to energetically 
favorable hydrolysis of ATP to ADP and inorganic phosphate \cite{theCell}. 

In a dynamical system with one-way coupling, an important notion is
 convective instability, where small perturbation is amplified to down-flow, although 
the system is linearly stable and provides a stationary state without such perturbation. 
We will show that the {\bf properties 1 and 2} are satisfied with this convective 
instability, as will be explained in the next section. It will be also shown that some of the 
one-way coupled dynamical system with convective instability can have input 
dependence. 

Here we consider a network of catalytic reaction elements, each of which is represented 
by differential equations showing excitatory dynamics. Each element is assumed to be 
connected uni-directionally. In other words, each chemical at the $(i-1)$-th site has an 
influence at the next $i$-th level. To be specific, we choose the model, 

\begin{equation}
\dot{\vec{X}^i} = \vec{F}(\vec{X}^i, \vec{X}^{i-1})\\
\label{model-0}
\end{equation}
with $\vec{X}^i = (x^i, y^i)$. 

In the present model, the input is represented just as a boundary 
condition $\vec{X}^0=const. $, while the response is given by the dynamics at the 
down-flow $i \gg 1$. We will find input dependence, or in other words boundary-condition sensitivity, and clarify its mechanism in the term of convective instability. 

Since a convectively unstable (CU) system is often sensitive to noise as described later, 
it is important to discuss how our signal transmission works in the presence of noise, 
due to chemical fluctuations. Indeed, it is shown that our mechanism works in the 
presence of noise, or rather it works best when the noise strength is of a medium size. 

Although our motivation is originated in signaling phenomena, our formulation of 
input-dependent dynamics is generally applied to any uni-directional chain of 
dynamical systems. In this respect, the mechanism of analogue and digital input-dependence, 
as well as our formula for their condition, will have general applicability. It 
may include neural networks\cite{Tsuda}, optical networks\cite{Ikeda-Otsuka}, open 
fluid flow, and other chain network of dynamical systems. 

The present paper is organized as follows. In \S 2, noise amplification by convective 
instability will be reviewed. In \S 3, a simple model using coupled chemical reactions is 
introduced for biological signaling pathway.  
In \S 4, input dependence in our model is presented, while its mechanism will be 
explained in \S 5 in connection with convective instability and its spatial change due to 
input signal. In \S 6, the relevance of the noise to the input dependence is described. It 
is shown that the input dependence is possible only within some range of noise 
amplitude. \S 7 is devoted to the discussion and the conclusion. 

\section{Convective instability in a one-way coupled system}

\subsection{convective instability}
\label{convective instability}

One-way coupled system is introduced\cite{KK0, Aranson} as an abstract model for an 
open flow system, for example, in a fluid system. By spatial amplification of disturbance 
at upper flow, the dynamics increases its complexity, as it goes to 
down-flow\cite{Aranson, KK-Crutch, Frederick-KK}. This change of dynamics is often 
triggered by convective instability. 

\begin{figure}[htbp]
\begin{center}
\includegraphics[scale=0.22]{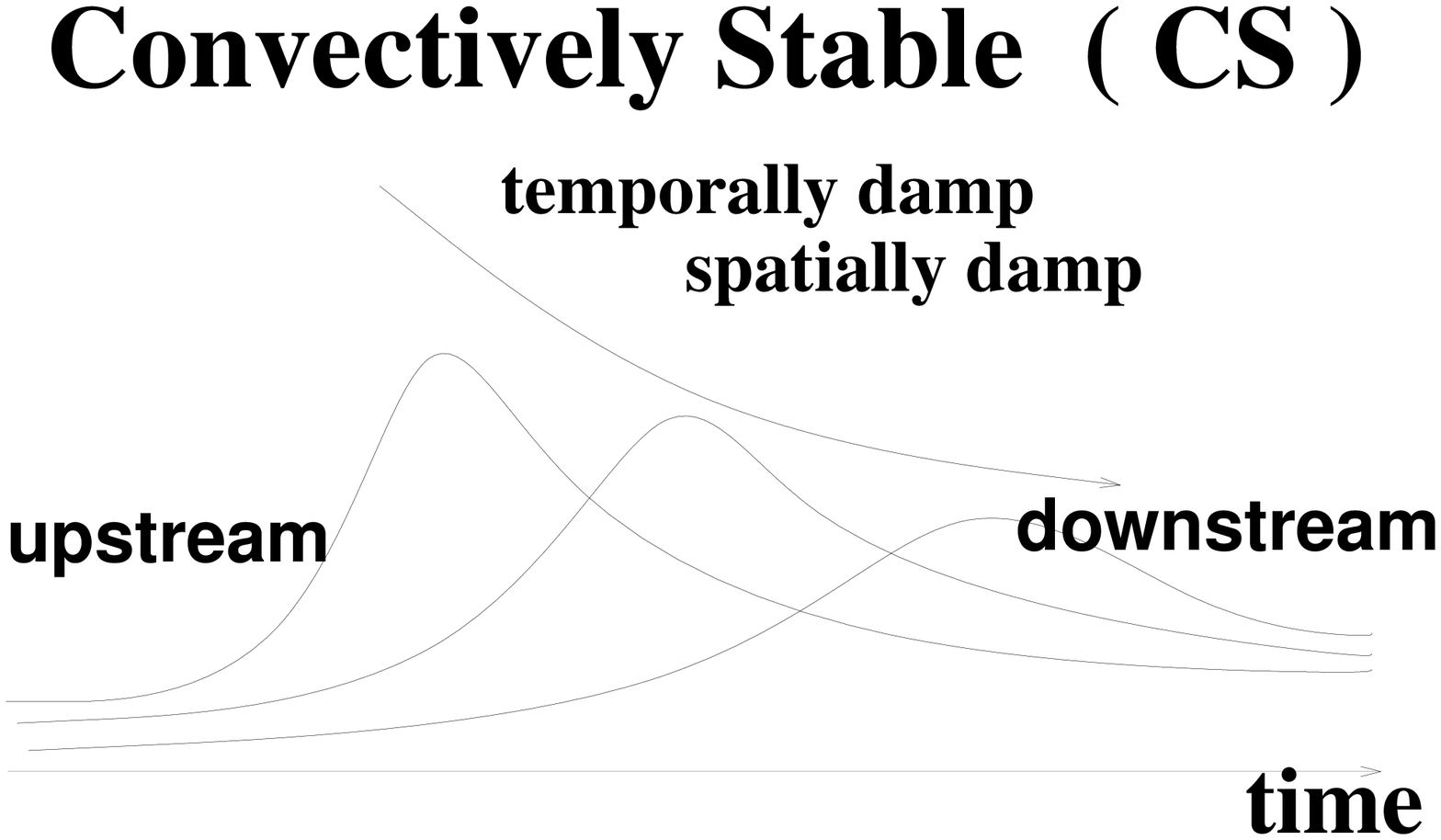}
\includegraphics[scale=0.22]{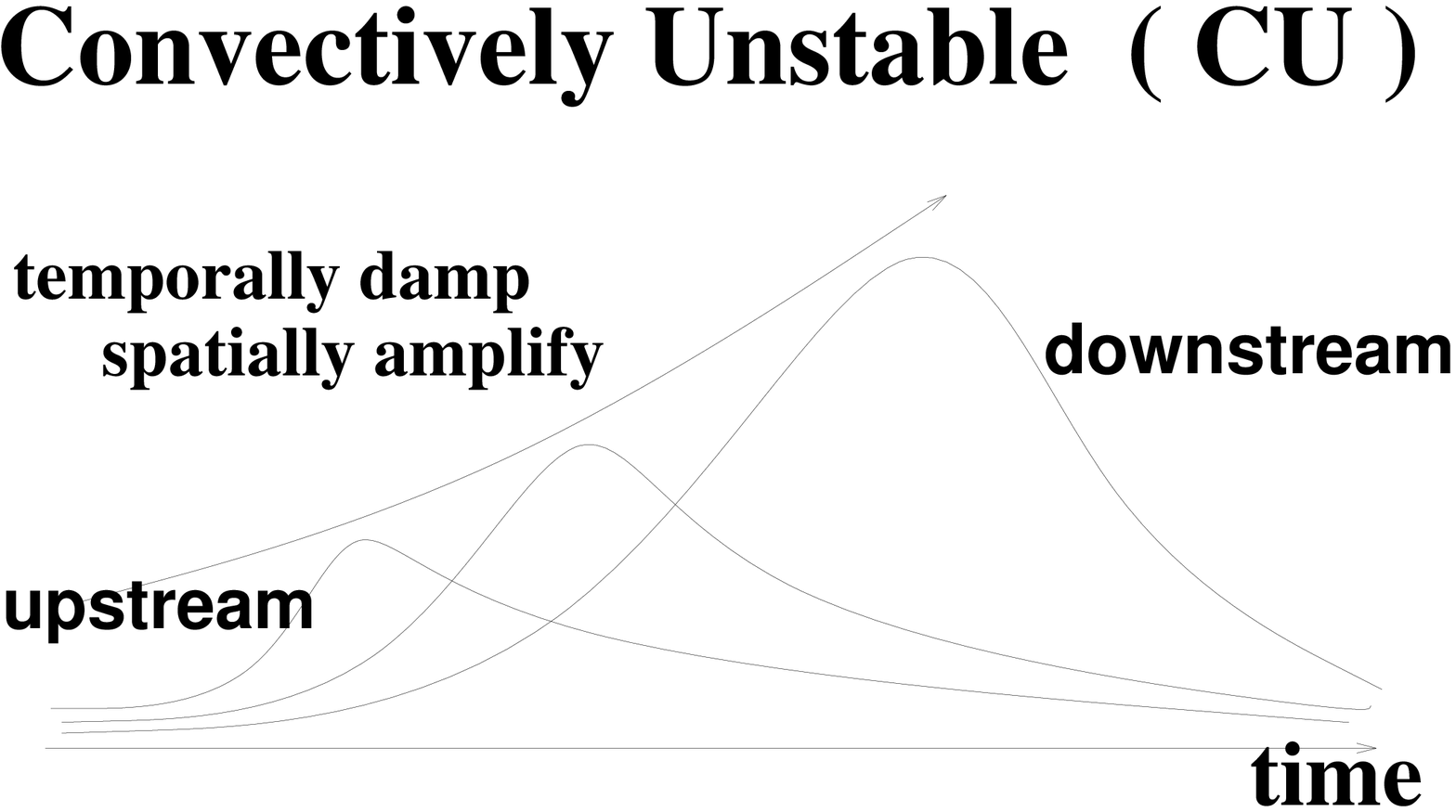}
\caption{Schematic representation of the evolution of perturbation}
\label{CU}
\end{center}
\end{figure}

Convective instability expresses how some perturbation is amplified along a flow. If a 
system is `convectively unstable' (CU), perturbation is spatially amplified and 
transmitted as in Fig.\ref{CU}. On the other hand, if a system is convectively stable 
(CS), perturbation at an upper flow is damped as it goes down-flow. Note that even if the 
perturbation is damped at each site (i.e., the system is linearly stable (LS) ), the system 
can be CU as shown in Fig.\ref{CU}.

Convective instability is quantitatively characterized by co-moving Lyapunov exponent 
$\lambda_v$, Lyapunov exponent observed from the inertial system moving with the 
velocity $v$ \cite{KK-Deissler, KK-Co-move}. For a given state, if 
$\mathop{max}_{v}\lambda_v$ is positive, the state is convectively unstable. The 
convective stability is supported by $\mathop{max}_{v}\lambda_v<0$, distinguishable 
from linear stability, given by the condition $\lambda_0<0$. The co-moving Lyapunov 
exponent is usually applied to an attractor, where chaos with convective instability is 
characterized by the positivity of $\mathop{max}_{v}\lambda_v$. It is however applied 
to any state, to characterize its stability. 

Since $\lambda_v$ characterizes the amplification of perturbation for the velocity $v$, 
the amplification per one lattice site is given by $\frac{\lambda _v}{v}$. Hence the 
amplification rate per one site is given by the spatial Lyapunov exponent 
$\lambda^S$\cite{Vulpiani};

\begin{equation}
\lambda^S = \mathop{max}_{v} \frac{\lambda_v}{v}
\end{equation}

\subsection{noise sustained structure in convectively unstable system}
\label{subsec:noise}
For a system with convective instability, noise plays an important role. When a system 
is linearly stable but convectively unstable, applied noise is spatially amplified and 
transmitted from up-flow to down-flow, until spatiotemporal structure is generated at 
the down-flow\cite{KK0, Deissler}. This structure is different from that for a system 
without noise. 

Mechanism of the structure formation is summarized as follows\cite{KK-Deissler, 
Deissler}. Assume that noise is added to LS but CU fixed point. Around the fixed point, 
the noise is spatially amplified and transmitted from up-flow to down-flow. The more it 
goes to down-flow, the larger oscillation can appear, until some stationary dynamics 
(such as periodic oscillation) is generated for $i \gg 1$. As long as the noise is added at 
the most up-flow element ($i = 0$), the down-flow dynamics remains same. 
This noise-induced structure in a convectively unstable system is a general feature in a one-way 
coupled system and important for our model. Of course, if the system is CS around the 
fixed points at all elements, noise is spatially damped, and no oscillation exists at the 
down-flow.

\section{Model}
As a simple signaling model, we choose a one-way coupled differential 
equations(OCDE)\footnote{OCDE is also studied by Aranson et.al\cite{Aranson}. }, 
where differential equation of each element can be regarded to express the dynamics of 
biochemical reaction. 

\subsection{dynamics of a single element}
\begin{figure}[hbtp]
\begin{center}
\includegraphics[scale=0.18]{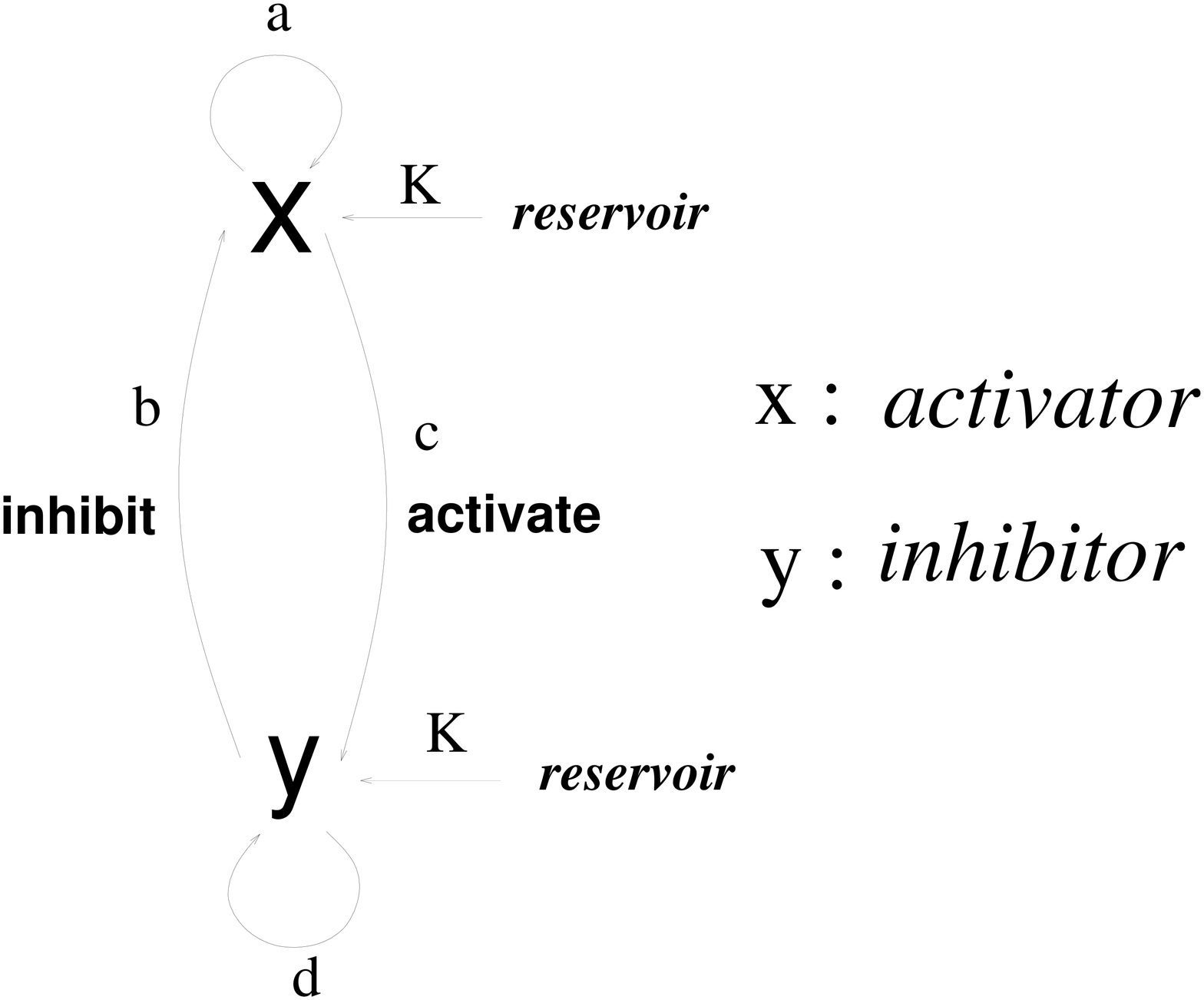}
\caption{ Schematic representation of biochemical reaction of a single element}
\label{BiochemicalReaction}
\end{center}
\end{figure}

As dynamics of a single element, we choose an auto-catalytic biochemical reaction 
system, consisting of activator $x$ and inhibitor $y$ (see Fig.\ref{BiochemicalReaction}). 
In other words, there is a set of two chemical variables for 
chemical concentrations, represented by differential equations with two degrees of 
freedom. 

As a specific example, we choose the following model, 

\begin{equation}
\left\{\begin{array}{ll}
\dot{x} = f(x, y)\\
\dot{y} = g(x, y)\
\end{array}\right. 
\end{equation}
with

\begin{equation}
\left\{\begin{array}{ll}
f(x, y) = x\left\{ (1 - x)ax - by \right\} + K\\
g(x, y) = y (cx - dy) + K \
\end{array}\right. 
\end{equation}
where the parameters $a, b, c, d$, and $K$ can be related with the rate constants of 
biochemical reaction. In the present paper, we set $K \ll 1$, $a, b, c, d \sim 1$ so that 
the system is satisfied with the three properties in \S1. In the present case, the 
suppression of the catalytic process is provided by the term $(1-x)$ (where $x$ is 
between 0 and 1), but other forms like Michael-Menten's can be adopted. Here details of 
the model are not important, and only the excitatory nature in the dynamics is 
necessary.

This single-element dynamics is chosen so that $(x , y)$ converges to a linearly stable 
fixed point $(x_*, y_*)$, where $f(x_*, y_*) = g(x_*, y_*) = 0$. There is neither a stable 
limit cycle nor another fixed point besides the above fixed point. 

\subsection{one-way coupling}
\label{susec:coupling}
Assuming that the identical set of 
dynamics is set at each $i$-th node, the coupling from each node to the next node is 
introduced as the activation process of the $i$-th node by the activator of the $(i-1)$-th 
node. In other words, we assume that the activator chemical at the $i$-th node is 
catalyzed by the $(i-1)$-th activator. This leads to the following one-way coupled 
differential equations. 

\begin{equation}
\left\{\begin{array}{ll}
\dot{x^i} & = f(x^i, y^i, x^{i-1})\\
\dot{y^i} & = g(x^i, y^i)
\end{array}\right. 
\label{eq:model}
\end{equation}

\begin{equation}
\left\{\begin{array}{ll}
f(x^i, y^i, x^{i-1}) & = x^i \left\{ (1 - x^i)(ax^i + \epsilon x^{i-1}) - by^i \right\} + K + 
{\eta}^i_x\\
g(x^i, y^i) & = y^i (cx^i - dy^i) + K + {{\eta}^i_y} \
\end{array}\right. 
\end{equation}
where $i$ denotes a spatial position or a level of signal pathway. Coupling constant 
$\epsilon$ corresponds to the rate constant of biochemical reaction between $(i-1)$-th 
activator and $i$-th activator, and $\eta$ represents white noise expressing 
concentration fluctuation of chemicals satisfying 

\begin{equation}
\langle {\eta}^j_p(t) {\eta}^k_q(t - \tau) \rangle_t = {\delta}_{p, q} {\delta}_{j, k} 
\delta (\tau) \sigma^2
\end{equation}
with $\sigma$, as the strength of the fluctuation, which is independent of $i$, $x$ or 
$y$. All the parameters are assumed to be independent of the node $i$, for simplicity. 
Again, the details of the coupling form are not important, as long as some nonlinear 
one-way coupling is included. 

This pathway transforms external signal into final response. Here, the input signal is 
given by the concentration $x^0$ (external signaling chemical) that appears as a 
parameter for the dynamics of $x^1$. This chemical concentration $x^0$ is set as a 
constant. The response is given by the concentration of chemicals at the down-flow 
($i \gg 1$), whose dependence on $x^0$ is studied. In other words, $x^0$ works as the 
boundary condition of the one-way coupling system. In the present paper, we do not 
consider its temporal change (like input oscillation), although its introduction will be an 
interesting problem in future. We study how the excitatory pulse appears at the down-flow. 

With regards to the connection to cell signaling problem, the node $i$ can be regarded 
either to one-dimensional spatial position or to a level of kinase reaction. In the latter 
case the use of identical reaction eq.(\ref{eq:model}) for each $i$ is of course, too drastic 
simplification. Still, it is useful to discuss a possible signaling mechanism in this simple 
case. Indeed the mechanism and concepts in the present paper can be straightforwardly 
applied even if the reaction equation differs by each element $i$.

\subsection{numerical results : pulse generation}
\begin{figure}[hbtp]
\begin{center}
\includegraphics[scale=0.5]{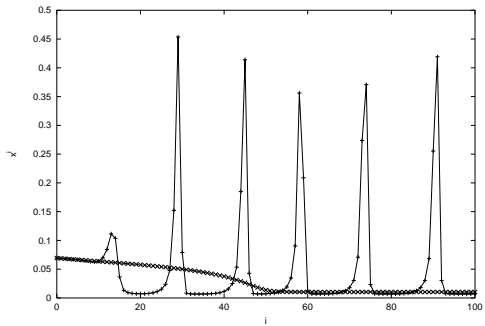}
\caption{Snapshot of space-amplitude plot for our system with noise ($+$ : 
$\sigma=10^{-4}$) and without noise ($\times$ : $\sigma=0$). With noise, the 
dynamics of each element at the down-flow is periodic. $a=0.4$, $b=5.12$, $c=2.0$, 
$d=3.55$, $K=0.0004$, $\epsilon=2.8$. Throughout the figures in the paper these 
parameter values are fixed.}
\label{Fig:Spatial plot}
\end{center}
\end{figure}

Since we have chosen our model to 
satisfy the linear stability of fixed points, all elements converge to a fixed point pattern, 
when the noise is not added as shown in Fig.\ref{Fig:Spatial plot}. The pattern shows 
dependence on the boundary at the up-flow, but it rapidly converges to a fixed value. 
The down-flow pattern is spatially constant whose value does not depend on the input. 
There is no stable attractor with oscillatory dynamics besides the above fixed point 
pattern. 

Here we are interested in the case that the fixed point pattern is convectively unstable 
at least some point. Then the mechanism mentioned at \S\ref{subsec:noise} works 
when some noise is added. We have found that a pulse-like oscillation is generated at 
the down-flow, as shown in Fig.\ref{Fig:Spatial plot} (see also Fig.\ref{fig:SpaceTimePlot}). 
Often, (i.e., for most parameter values and input values), the 
generated oscillation is periodic. At the down-flow, the oscillation becomes periodic both 
in space and time. In other words, the pulse is transmitted to down-flow without 
changing its pattern (i.e., amplitude or frequency). Note that the limit cycle has to be 
CS to be transmitted to down-flow without being changed by noise added at each 
element. Still, this limit cycle disappears if the noise is turned off. 

\section{Input dependence}
To see the input dependence of the pulse-like solution, we 
have changed the concentration $x^0$ and studied how the dynamics at the down-flow 
changes. Here the boundary (input) concentration is taken to be fixed in time, and 
temporal information of the input is discarded. 

Two types of the input-dependence are discovered, as to the down-flow dynamics, given 
by the behavior of $(x^i(t), y^i(t))$ at $i \gg 1$. The first one is a `digital' (i.e., 
threshold-type) change, represented as a different phase of dynamics, such as fixed 
points, oscillation, and so forth, while the second one is `analogue' change of the 
frequency or amplitude of the oscillation. Note that these input dependence of the 
down-flow dynamics are kept even for $i \rightarrow \infty$ where the dynamics 
spatially converges. 

\subsection{digital change}
\begin{figure}[hbtp]
\begin{center}
(a)\includegraphics[scale=0.45]{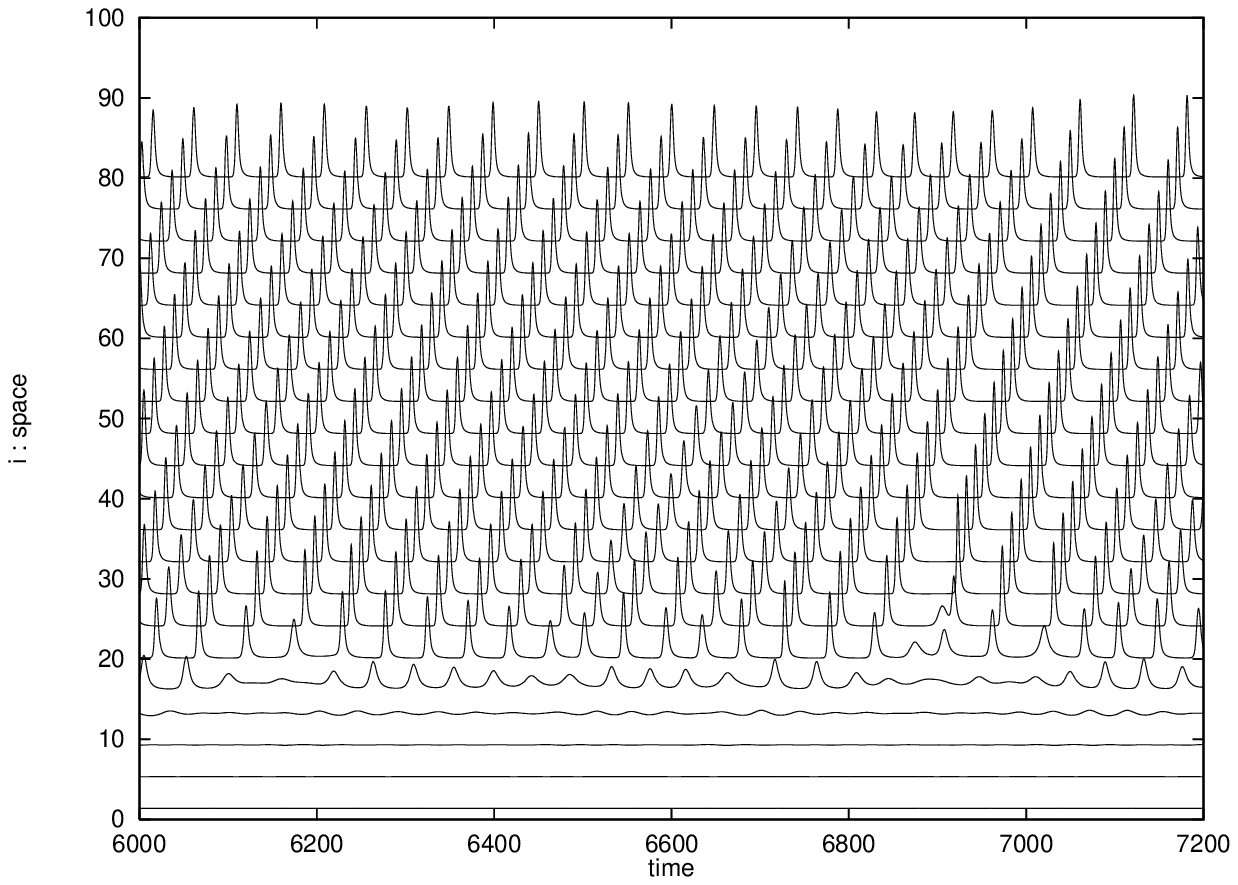}
(b)\includegraphics[scale=0.45]{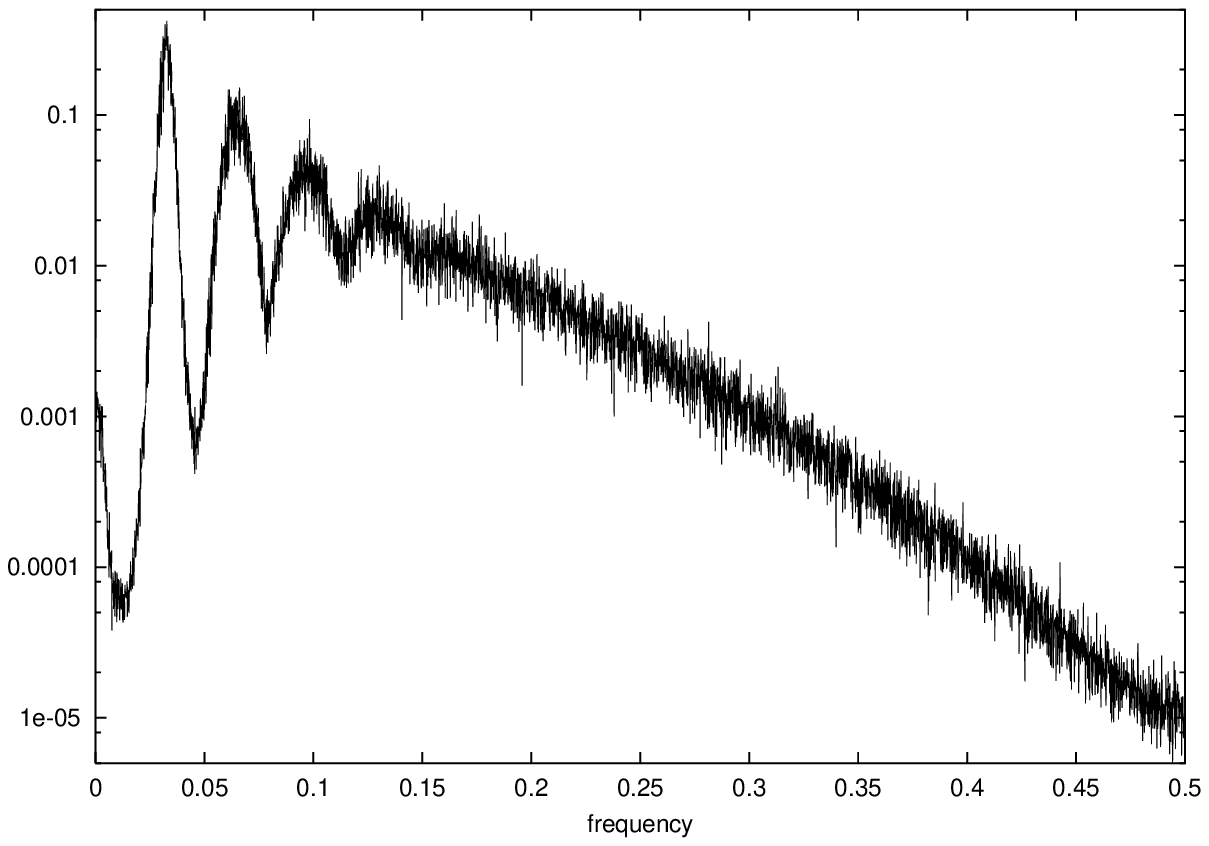}
(c)\includegraphics[scale=0.45]{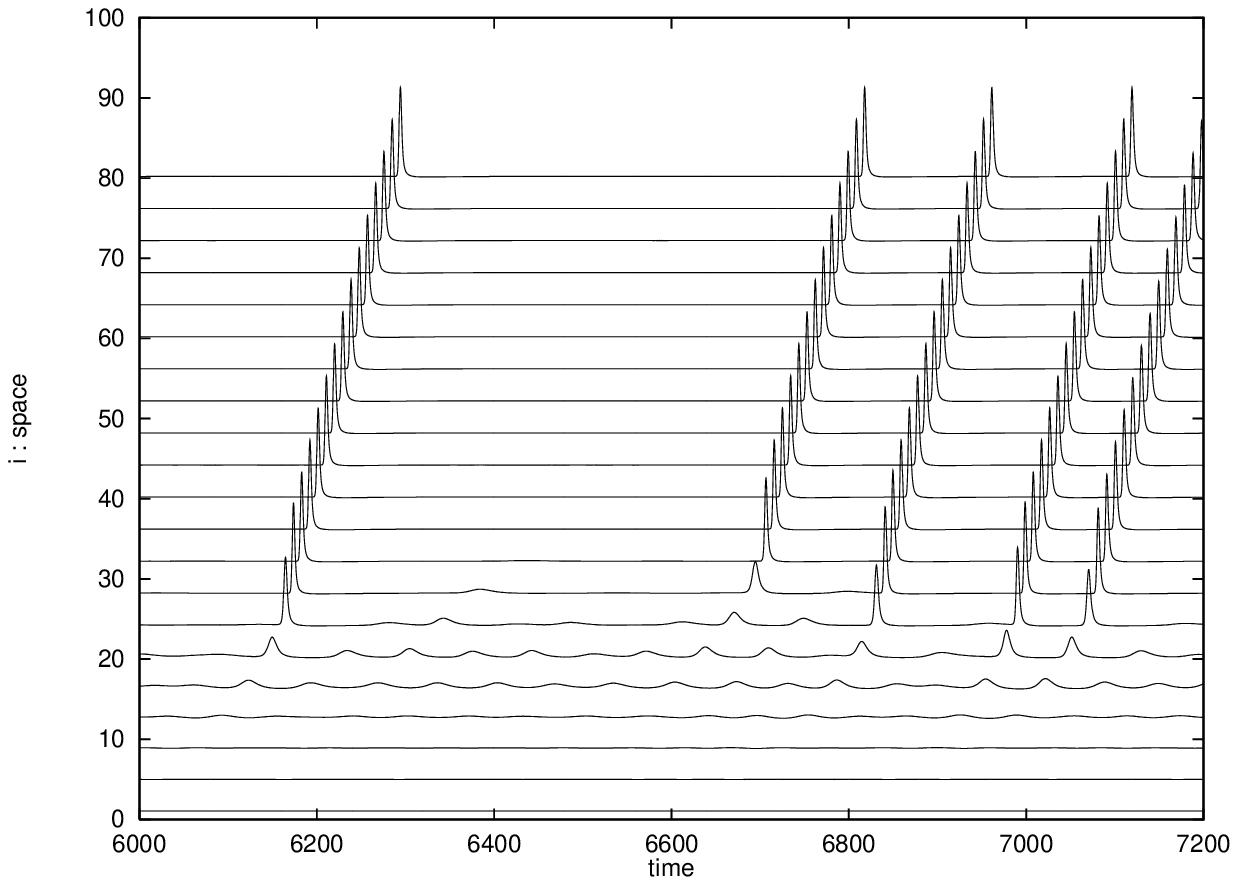}
(d)\includegraphics[scale=0.45]{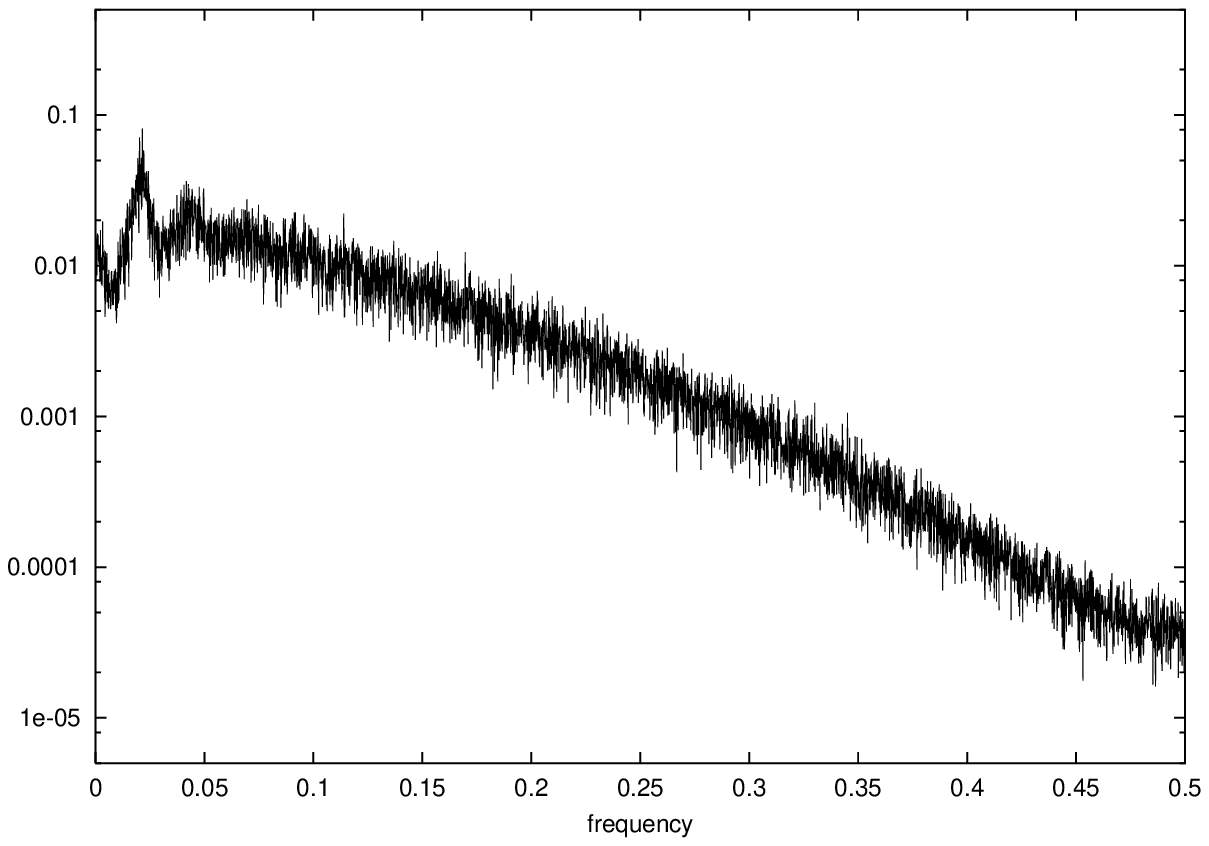}
\caption{Spatiotemporal plot of $x^i(t)$ (a), (c) and the corresponding power spectrum 
(b), (d) with $i=80$. The input value $x^0$ are $0.070$ (a, b) and $0.054$ (c, d). (a) and 
(b) show periodic oscillation at down-flow, (c) and (d) give stochastic oscillation. $\sigma 
= 1.0 \times 10^{-4}$. Parameters $a, b, c, d, K, \epsilon$ are same as Fig.3. }
\label{fig:SpaceTimePlot}
\end{center}
\end{figure}

With the application of noise to all elements, the response shows the following three 
phases successively, with the increase of the input value (the boundary condition) $x^0$. 
The difference of the phases is shown in Fig.\ref{fig:SpaceTimePlot} with the 
spatiotemporal pattern and the corresponding power spectrum for the time series of 
$x^i$ at the down-flow. 

\begin{figure}[hbtp]
\begin{center}
\includegraphics[scale=0.45]{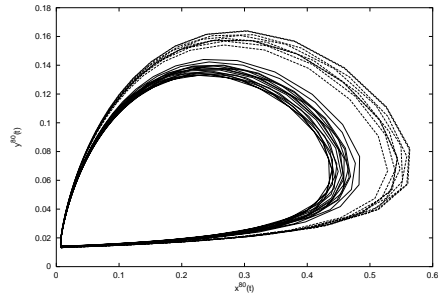}
\caption{Phase space at $i=80$ to different input value $x^0$. 
 Parameters $a, b, c, d, K, \epsilon$ and intensity of noise $\sigma$ are same as Fig.4. 
The dotted line corresponds to Fig.4(c) (s-phase), 
while the solid line corresponds to Fig.4(a) (p-phase). }
\label{fig:phasespace}
\end{center}
\end{figure}

\begin{description}
\item[periodic oscillation phase (p-phase)] (limit cycle)

At the down-flow, periodic oscillation with a large amplitude is generated, as shown in 
Fig.\ref{fig:SpaceTimePlot}(a). The motion at the down-flow is quite regular, in the 
presence of noise, as is shown in the power spectrum of $x^i$ at $i=80$ (see Fig.\ref{fig:SpaceTimePlot}(b)). 
The orbit of $(x^i, y^i)$ is plotted in the phase space in Fig.\ref{fig:phasespace}. 
Note that the orbit does not pass through the fixed point of the 
noiseless case. 

As mentioned at \S\ref{susec:coupling}, if the noise is turned off, the oscillation damps 
and each element dynamics converges to a fixed point. On the other hand, the 
oscillation is CS and transmitted without influenced by noise. 

\item[stochastic oscillation phase (s-phase)]

At the down-flow, stochastic oscillation with a large amplitude is generated. The 
interval between the peaks is stochastic and longer than the period of oscillation in the 
p-phase (see Fig.\ref{fig:SpaceTimePlot}(c) for the spatiotemporal pattern). 
Amplification and transmission of noise are regular at the p-phase while they are 
possible only stochastically at the present phase. Peaks in the power spectrum of the 
down-flow dynamics are no longer observed, as shown in Fig.\ref{fig:SpaceTimePlot}(d), 
and are replaced by the broad band spectra. The orbit stays close to the same fixed point 
of the noiseless case for long time, emits a pulse train stochastically once, and returns to 
the former fixed point. (See Fig.\ref{fig:phasespace} where the orbit of $(x^i, y^i)$. )

Although the oscillation is aperiodic, each pulse itself remains to be CS and is 
transmitted without influenced by noise. 

\item[fixed point phase (f-phase)]

No pulse with large amplitude is generated. The orbit stays around the fixed points 
with some fluctuation ($\sim \sigma$). No spatial amplification is observed. At the 
transition from the s-phase to the present phase, the frequency of the pulse generation 
goes to zero. 

\end{description}

We have measured the distribution of the interval between the peaks of $x^i$ at the 
down-flow, as the boundary $x^0$ is changed. At the p-phase, the distribution has a 
sharp peak, while no typical peaks are observed at the distributions for the s-phase.

\begin{figure}[hbtp]
\begin{center}
\includegraphics[scale=0.5]{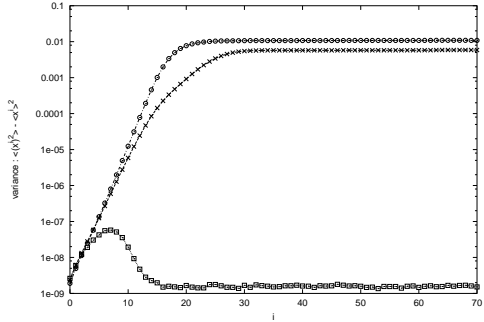}
\caption{Dependence of the variance $<(x^i)^2>-<x^i>^2$ on the site $i$. The top one 
($\circ$) corresponds to the input of Fig.4(a) (p-phase), the middle one ($\times$) to 
Fig.4(c) (s-phase), while the bottom one ($\Box$) corresponds to the fixed point phase 
(f-phase). The dynamics spatially converges at $i \sim 30$ and input dependence is 
kept for $i \rightarrow \infty$. Parameters $a, b, c, d, K, \epsilon$ and intensity of 
noise $\sigma$ are same as Fig.4. }
\label{fig:Bunsan}
\end{center}
\end{figure}

To see if the dynamics converges at a down-flow, we have measured the variance of $x$, 
that is $<(x^i)^2>-<x^i>^2$, with $<. . >$ as the temporal average. 
Fig.\ref{fig:Bunsan} gives the spatial change of the variance for three different boundary 
conditions $x^0$. 
The upper plot corresponds to the p-phase, the middle to the s-phase, 
and the bottom corresponds to the f-phase. As in the figure, the dynamics spatially 
converges at $i \sim 30$, while the input dependence is preserved for $i \rightarrow 
\infty$. (Indeed we have confirmed that the variance $<(x^i)^2>-<x^i>^2$ remains 
different for each boundary condition.)
All of our results show that the input 
dependence is not a transient in space but is kept for $i \rightarrow \infty$. 

\subsection{analogue change}
\begin{figure}[hbtp]
\begin{center} 
\includegraphics[scale=0.7]{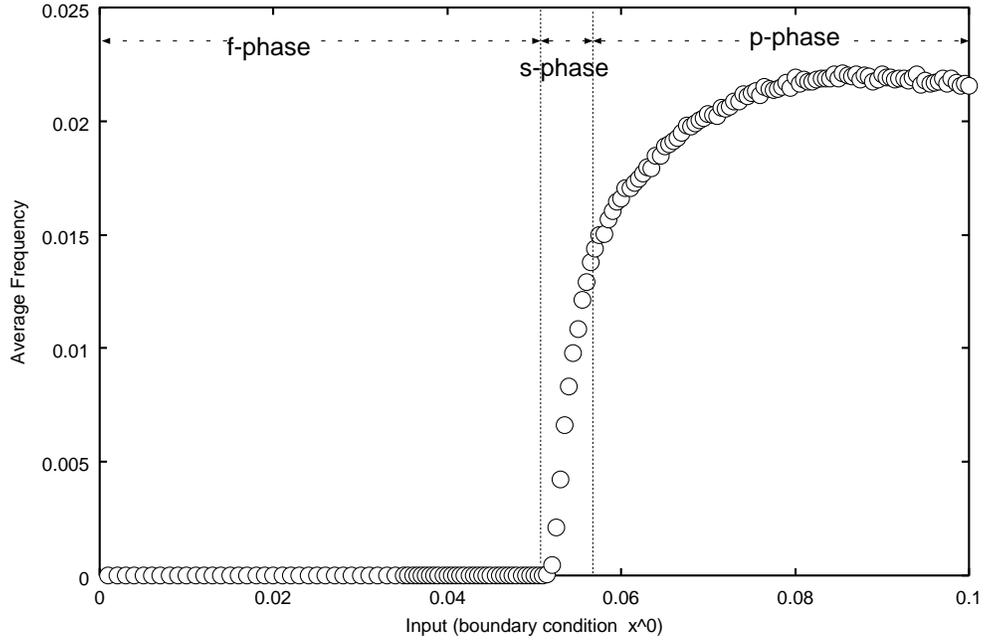}
\caption{Input dependence of the average frequency of pulses at down-flow is plotted. 
From the power spectrum, it is shown that the p-phase lies at $x^0 > 0.054$, and s-
phase at $0.054 > x^0 > 0.051$. For both the phases, the average frequency changes 
with the input value $x^0$. Parameters $a, b, c, d, K, \epsilon$ and intensity of noise 
$\sigma$ are same as Fig.4. }
\label{fig:power-InputDep}
\end{center}
\end{figure}

Analogue dependence on input is given by a continuous change of the frequency and 
amplitude of the oscillation (at $i \gg 1$) with the input, i.e, the boundary condition 
$x^0$. It is observed both in the periodic and stochastic oscillation phases. 
Fig.\ref{fig:power-InputDep} shows input dependence of the average frequency of the 
down-flow dynamics. In both the phases, the average frequency of 
the down-flow dynamics changes continuously with the boundary value $x^0$. The 
down-flow dynamics shows not only digital but also analogue dependence on the 
boundary. We have also measured power spectrum of the down-flow dynamics. The 
position of the peak frequency (accordingly the response dynamics) continuously 
changes with the input $x^0$. 

\section{Mechanism}
\label{sec:Mechanism}
How is the input dependence formed? Recall the following two properties described in 
\S\ref{subsec:noise}: First, the formation of a pulse train structure depends on if the 
fixed point solution is CS or CU. Second, the nature of the generated oscillation is 
influenced by the degree of convective instability of the fixed point at the region where 
the pulse is being formed. 

By referring to these properties, the mechanism of input dependent dynamics is 
sketched as follows: Depending on the input, the fixed point values at the upper flow 
change, which leads to the difference in convective instability. Accordingly the degree of 
the noise amplification changes, and the generated dynamics at the down-flow is 
different. In this section we study the above mechanism of the input dependence in the 
term of the convective instability of the fixed point. 

Without noise, all elements converge to fixed points
 $(x^i_*, y^i_*)$ that depend on the site $i$. 
Now we introduce the site-dependent co-moving Lyapunov exponent 
$\lambda_v(i)$ and spatial instability exponent $\lambda^S(i)$ of $(x^i_*, y^i_*)$. 
We will see that the input dependence can be explained by spatial change of 
$\lambda^S(i)$ for $(x^i_*, y^i_*)$. 

\subsection{spatial convergence of fixed point}
\label{subsec:Spatial convergence}
In our model, $(x^i_*, y^i_*)$ is determined as follows. 

\begin{equation}
\left\{\begin{array}{ll}
\dot{x^i_*} = f(x^i_*, y^i_*, x^{i-1}_*) & =0\\
\dot{y^i_*} = g(x^i_*, y^i_*) & =0\
\end{array}\right. 
\label{eqn:recursion}
\end{equation}

\begin{figure}[hbtp]
\begin{center}
\includegraphics[scale=0.8]{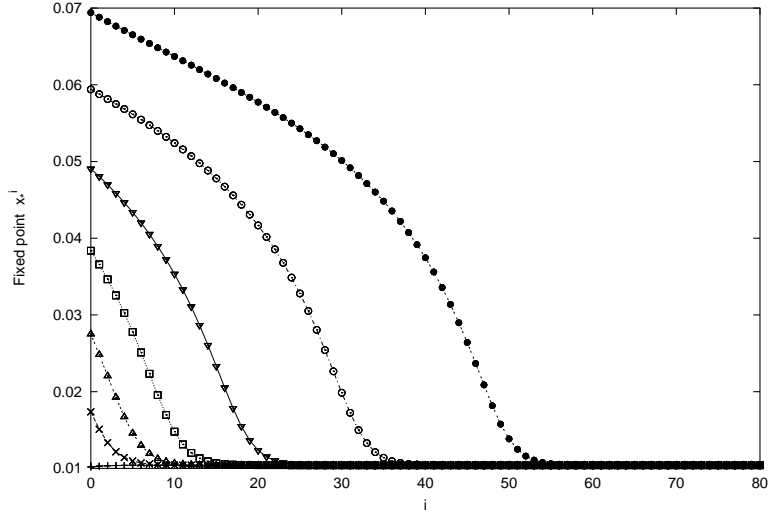}
\caption{The fixed point $x_*(i)$ is plotted for different input values $x^0$. 
Parameters $a, b, c, d, K, \epsilon$ are same as Fig.4. Different marks corresponds to 
different input values, $x^0 = 0.01(+)$, $0.02(\times)$, $0.03(\bigtriangleup)$,
$0.04(\Box)$, $0.05(\bigtriangledown)$, $0.06(\circ)$, and $0.07(\bullet)$. }
\label{Fig.fixpt}
\end{center}
\end{figure}

Eq.(\ref{eqn:recursion}) gives a recursion equation of from $(x^{i-1}_*, y^{i-1}_*)$ to 
$(x^i_*, y^i_*)$. Given $x^0$ at the left boundary, this equation can be solved 
successively for $i = 1, 2, . . . . $ by the iteration. 
The trace of spatial convergence, $(x^1_*, y^1_*)$ $\rightarrow$ $(x^i_*, 
y^i_*)$ $\rightarrow$ $(x^{\infty}_*, y^{\infty}_*)$ is decided as in Fig.\ref{Fig.fixpt}. 
In our model, $(x^{\infty}_*, y^{\infty}_*)$ is decided uniquely\footnote{It is an 
interesting future problem to study a case with a multiple solution of 
$(x^{\infty}_*, y^{\infty}_*)$. } to $\forall x^0$. 
We use the notation, 
\begin{equation}
\begin{array}{rcl}
(x_*, y_*) & \equiv & (x^{\infty}_*, y^{\infty}_*) \\
\lambda^S_* & \equiv & \lambda^S_*(\infty)
\end{array}
\end{equation}

Now let us study the convective stability of this fixed point. Since the fixed point is site 
dependent, we extend the co-moving Lyapunov exponent to a spatially local one, by 
measuring the amplification of perturbation to the next site from each site for a given 
velocity. Technically this amplification rate is measured as the co-moving Lyapunov 
exponent for a snapshot homogeneous pattern $(x^l, y^l) = (x^i_*, y^i_*)$ for $l \geq 
i$ (see Appendix). 

\begin{figure}[hbtp]
\begin{center}
(a)\includegraphics[scale=0.49]{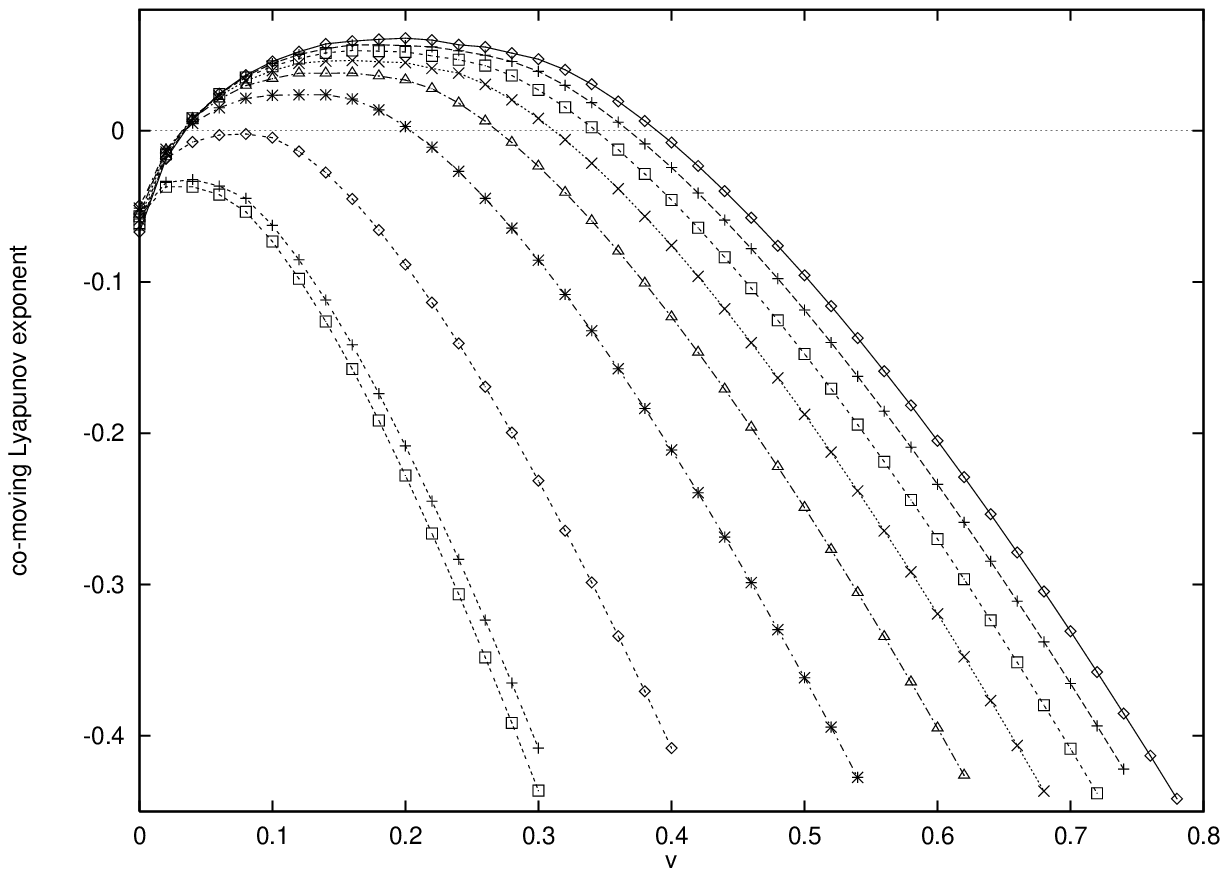}
(b)\includegraphics[scale=0.49]{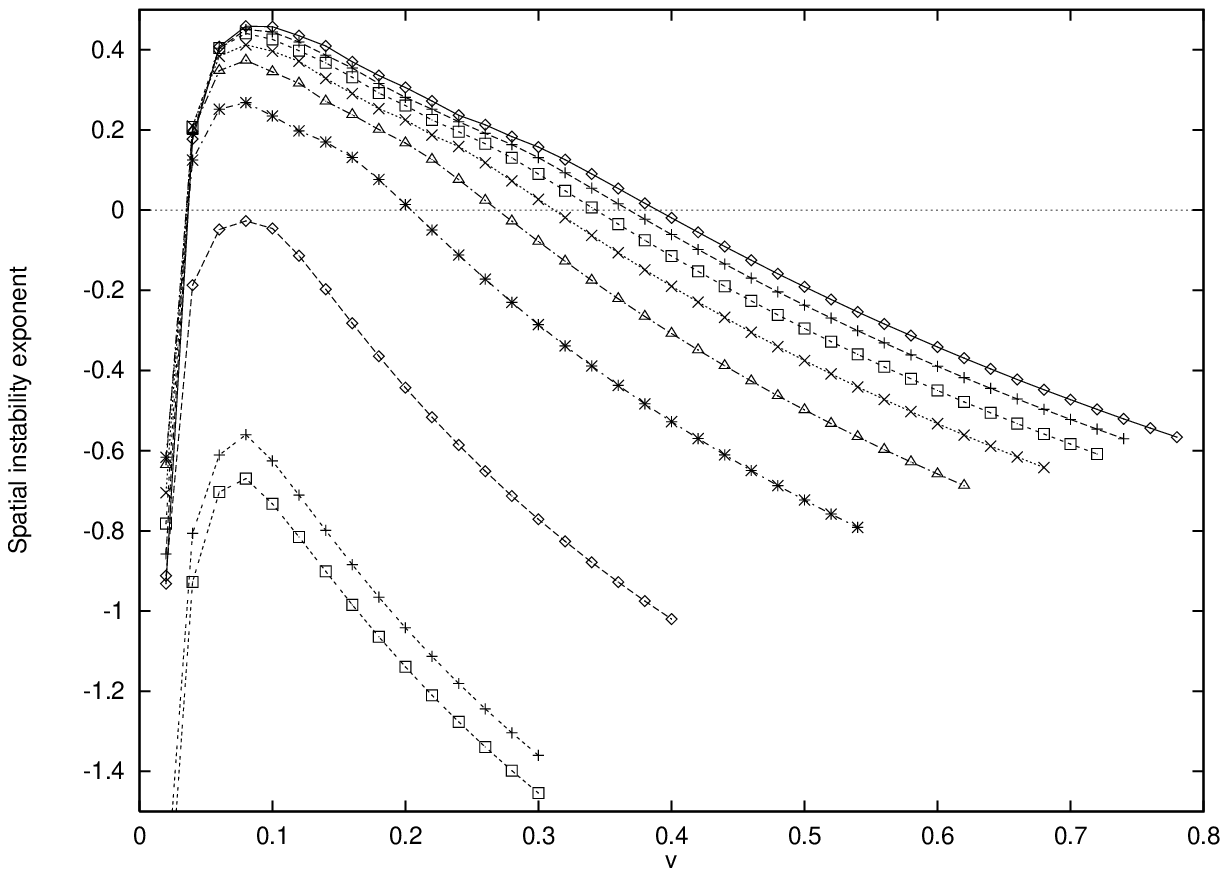}
\caption{Co-moving Lyapunov exponents $\lambda_v(i)$ (a) and the spatial instability 
exponents $\frac{\lambda_v(i)}{v}$ (b) are plotted as a function of $v$, for 
$i=$ $0(\Diamond)$, $5(+)$, $10(\Box)$, $15(\times)$, $20(\bigtriangleup)$, 
$25(\ast)$, $30(\Diamond)$, $35(+)$, $40(\Box)$. Parameters $a, b, c, d, K, 
\epsilon$ are same as Fig.4. The boundary condition is given by $x^0 = 0.060$. 
$\frac{\lambda_v(i)}{v}$ is maximum at $v \sim 0.1$ for every site. }
\label{Fig.ram_v}
\end{center}
\end{figure}

Fig.\ref{Fig.ram_v}(a) gives the co-moving Lyapunov exponent $\lambda_v(i)$ of the 
fixed point at each site, while $\frac{\lambda_v(i)}{v}$ of the fixed point at each site is 
given in Fig.\ref{Fig.ram_v}(b). Here, site dependence is clearly seen. At $i \leq 30 
$ the fixed point is CU, since $\lambda_v(i)$ is positive at $v \sim 0.1 $. On the other 
hand, at $i \geq 30 $, the fixed point is CS, $\lambda_v(i)$ is negative for all $v$. In 
our model, the fixed points change from convectively unstable to stable ones, as the site 
goes down-flow.

\begin{figure}[hbtp]
\begin{center}
\includegraphics{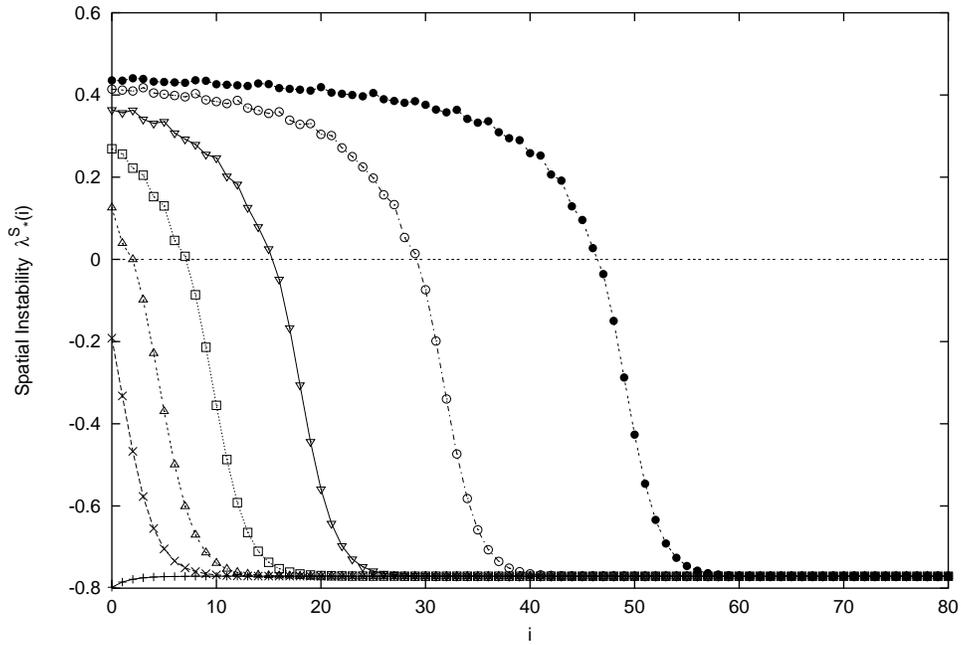}
\caption{The spatial instability exponent $\lambda^S_*(i)$ of the fixed point. 
Different marks represent different input value, $x^0 = 0.01(+)$, $0.02(\times)$, 
$0.03(\bigtriangleup)$, $0.04(\Box)$, $0.05(\bigtriangledown)$, $0.06(\circ)$, 
and $0.07(\bullet)$, corresponding to Fig.8. }
\label{Fig.ram_S}
\end{center}
\end{figure}

Recall the growth rate of the perturbation to the next sites is given by the spatial 
instability, $\lambda^S(i) = \mathop{max}_{v} \frac{\lambda_v(i)}{v}$. In our model, 
such velocity $v$ that maximizes $\frac{\lambda_v(i)}{v}$ is independent of $i$, as is 
shown in Fig.\ref{Fig.ram_v}(b). Change of $\lambda^S(i)$ with the input $x^0$ is 
shown in Fig.\ref{Fig.ram_S}. In our model, the change is monotonic with respect to 
site $i$ \footnote{It is interesting to study the case with non-monotonic convergence of
 fixed points (such as a spatial periodic or chaotic case), in future. }, 
either by decreasing from $\lambda^S_*(1)$ to $\lambda^S_*$ or increasing. 

Now we discuss the relationship between $\lambda^S_*(i)$ and input dependence. In 
Table \ref{tab:type}, we have classified our dynamics into three types according to the 
convective stability at the up-flow and down-flow. 

\begin{table}[hbtp]
\caption{Composition of CS or CU states at up-flow and down-flow}
\label{tab:type}
\begin{center}
\begin{tabular}{|c|c|c|c|} 
\hline
  & fixed point & fixed point & limit cycle \\
  & at up-flow & at down-flow & at down-flow \\
\hline Type 1 & CS   & CS   & none   \\
\hline Type 2 & CU   & CS   & CS   \\
\hline Type 3 & CU or CS & CU   & CS   \\
\hline
\end{tabular}
\end{center}
\end{table}

First, for the Type1, no oscillation is generated, since the fixed point is CS. The fixed 
point approaches a constant value for $i \rightarrow \infty$, and the state value at the 
down-flow is input independent. For the Type3 the fixed point at the down-flow is CU, 
and spatial amplification from a fixed point always occurs, independently of the input 
value $x^0$. Hence, no digital change occurs. 
Indeed, in numerical simulations, the 
digital change from a fixed point to oscillation is observed 
only at the Type2 case only. In this case, there are two CS states at $i \rightarrow 
\infty$. One is the fixed point $(x_*, y_*)$ which is LS (Linearly Stable), and the other 
is a limit cycle. 
According to the nature of the up-flow dynamics, either of the two 
states is selected, that leads to a digital change in the dynamics. 

\subsection{mechanism of digital change}
\label{Mechanism of digital change}
In this 
section mechanism of digital change is quantitatively analyzed according to the co-
moving Lyapunov exponent. 

Since the fixed point becomes convectively stable at the down-flow, the oscillatory 
dynamics should be formed before the fixed point is stabilized. Hence the condition for 
the formation of oscillatory dynamics is given by the competition between two time 
scales, the scale $i_u$ where the lattice remains to be CU, and other the scale $i_g$, 
required to the generation of a pulse, i.e. the ``growth scale'' for perturbation. The scale 
$i_u$ is given by the site where the fixed point $(x^i_*, y^i_*)$ becomes CS, and it is 
given by the condition $\lambda^S(i) > 0$ at $i < i_u$ and $\lambda^S(i) < 0$ at $i > i_u$. 

The scale $i_g$ is estimated as the scale where the noise is amplified to the scale of O(1), 
to generate oscillatory dynamics. As long as the noise is small, the amplification rate of 
noise is determined by the spatial instability $\lambda^S_*(i)$ of fixed point $(x^i_*, 
y^i_*)$. Hence $i_g$ is roughly estimated as follows. 

\begin{equation}
\sigma \prod^{i_g}_{i=0}e^{\lambda^S_*(i)} = \sigma 
e^{\sum^{i_g}_{i=0}\lambda^S_*(i)} \sim 1
\label{eq:condition}
\end{equation}

\begin{equation}
\sum^{i_g}_{i=0}\lambda^S_*(i) \sim log \frac{1}{\sigma}
\label{eq:i_g}
\end{equation}

\begin{figure}[hbtp]
\begin{center}
\includegraphics[scale=0.75]{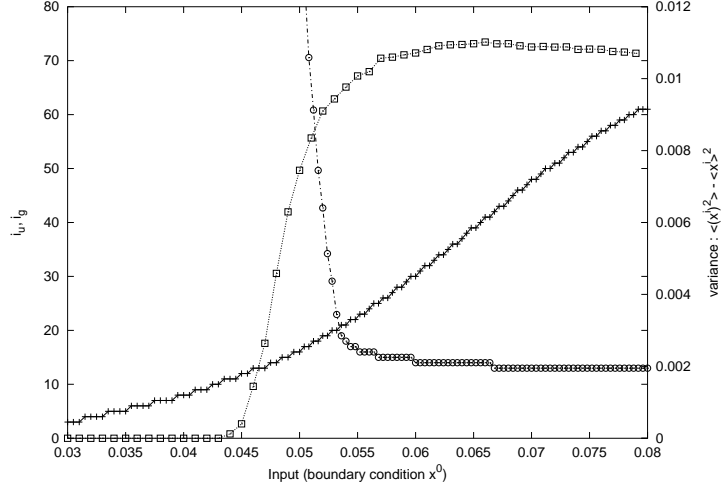}
\caption{Dependence of $i_u$ ($+$) and $i_g$ ($\circ$) on input value $x^0$. For the 
reference the change of variance of $x^{80}$ ($\Box$) is plotted. Parameters $a, b, c, d, 
K, \epsilon$ are same as Fig.4. Intensity of noise $\sigma = 0.001$. }
\label{Fig.InputDepOni_ui_g}
\end{center}

\end{figure}

Since the perturbation is no longer amplified at a CS state, it is necessary that noise is 
amplified while the dynamics at the site remains to be CU. The noise has to be 
amplified to O(1) before $i \sim i_u$, the condition $i_g \stackrel{<}{\sim} i_u$ is 
imposed to have oscillatory dynamics. As the input (boundary) changes, fixed points 
$(x^i_*, y^i_*)$ change accordingly, and both $i_u$ and $i_g$ change. Then the relation 
between $i_g$ and $i_u$ can change qualitatively. Fig.\ref{Fig.InputDepOni_ui_g} 
shows input dependence of $i_g$ and $i_u$ for our model for parameters belonging to 
Type2. Note that the relationship $i_g \stackrel{>}{<} i_u$ changes with the input 
$x_0$. For $x^0 > 0.065$, $i_g$ is smaller than $i_u$, and indeed we have observed 
oscillatory dynamics with a large amplitude. Around $x^0 \sim 0. 065$, $i_g \sim i_u$, 
where stochastic oscillation is found with intermittent pulse generation from the fixed 
point state. No pulse is generated for $x^0 < 0.065$. 

We can summarize the change of dynamics according to the relationship between $i_g$ 
and $i_u$, as follows:

\begin{itemize}

\item $i_g < i_u$ : This gives {\bf periodic phase}, where stationary oscillation is formed 
at down-flow. Since the generated oscillation itself is CS, the pulse is transmitted to 
down-flow without affected by noise. As long as noise is added stationarily (at least at 
the up-flow), spatially stationary state is formed at the down-flow. 

\item $i_g \sim i_u$ : This corresponds to the {\bf stochastic phase}, where the 
formation of pulse is strongly sensitive to noise at the moment. Formed pulse itself is 
CS and can be transmitted to down-flow stationarily, but its formation is intermittent. 
Indeed, we have measured $\frac{i_u}{i_g}$ at the parameters of the stochastic phase, 
and the value stays around $1.0 \sim 1.2$. 

\item $i_g = \infty$ ($i_g > i_u$) : This give a {\bf fixed point phase}, since the fixed 
point becomes CS before a pulse is generated. No pulse exists, and the down-flow 
dynamics remains around the fixed point with some noise. 
Indeed, the condition(\ref{eq:condition}) is not satisfied for any $i_g$. 

\end{itemize}

\begin{figure}[hbtp]
\begin{center}
\includegraphics[scale=0.4]{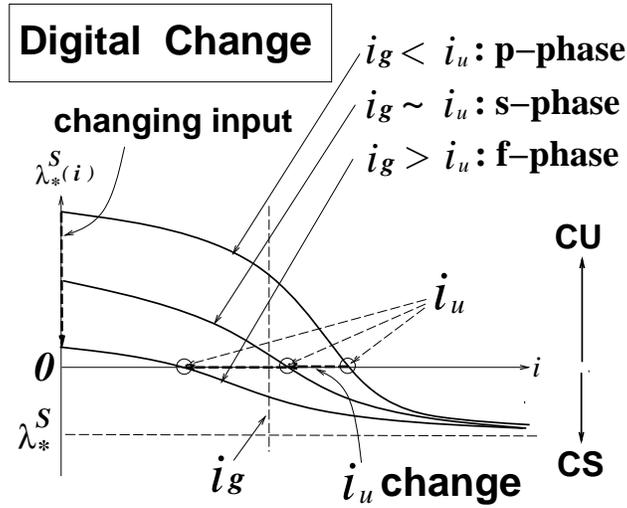}
\caption{Schematic diagram of the change of 
$\lambda^S_*$, corresponding to each of the three phases. }
\label{fig:DigitalChange}
\end{center}
\end{figure}

This summarizes how the digital change appears for our dynamics. It should be noted 
that the above mechanism generally holds in a one-way coupled system with the fixed 
points convectively unstable at the up-flow and stable at the down-flow. Now it is also 
clearer why the digital change is seen only for the Type2. For the Type1, $i_u$ is 
$\infty$, and only the fixed point phase appears, while for the Type3, $i_u$ is zero, and 
the pulse is formed irrespective of the boundary condition (input). The mechanism is 
shown schematically in Fig.\ref{fig:DigitalChange}. 


\subsection{mechanism of analogue change}
\label{Mechanism of analogue change}

Here we discuss the analogue change against input, that 'transforms' the input 
concentration to the response frequency. The rate of formation of pulse is expected to 
depend on the fixed points and the convective instability before the pulse train is formed. 
On the other hand, as the site goes down-flow, the fixed point approaches $(x_*, 
y_*)$ independent of the input $x^0$. In other words, the spatial instability is 
independent of the input, after the fixed point converges closely to $(x_*, y_*)$. Hence, 
only if the scale $i_g$, required to form the pulse, is smaller than the scale of the above 
convergence, the dynamics generated at the down-flow is expected to depend on the 
input. 

As a simple measure of the relaxation length $i_r$ of fixed points, we will introduce a 
'half-decay' scale. 
Since the spatial convergence of $\lambda^S_*(i)$ is monotonic as 
described in \S\ref{subsec:Spatial convergence} and $\lambda^S_*(i)$ characterizes 
the amplification of noise, we use the convergence of $\lambda^S_*(i)$ to define $i_r$. 
As the half-decay scale of $\lambda^S_*(i)$, $i_r$ is defined such site that 
satisfies $\lambda^S_*(i_r) = \frac{\lambda^S_* +\lambda^S_*(0) }{2}$. Roughly 
speaking, $\sum_j^i \lambda^S_*(i)$ changes sensitively on $i$ for $i < i_r$, while for 
$i > i_r$, it weakly depends on $i$. This is expected, for example, by assuming that 
$\lambda^S_*(i)$ exponentially relaxes to $\lambda^S_*$. 

Now let us focus on the relationship between $i_g$ and $i_r$. If the convergence scale 
$i_r$ is much smaller than $i_g$, the rate of amplification of noise given by
$\sum_j^i \lambda^S_*(j)$ depends little on the input. 
On the other hand, if $i_r$ is larger than 
$i_g$, the amplification rate largely depends on the input. Since this change of 
amplification rate is continuous with the input value, analogue change of response is 
expected. Hence the following two cases are classified. 

\begin{description}\item[Case A] : $i_g < i_r$ 

At the region where the pulse is formed ($i < i_g$), the spatial convergence of fixed 
points is not yet completed. The noise amplification rate there is dependent on the input. 
Thus the nature of generated wave at $i \sim i_g$ depends on the input. Since the 
formed wave pattern is convectively stable, this dependence on the input is preserved to 
down-flow. 

\item[Case B] : $i_g > i_r$ 

Since the fixed point has almost converged to $(x_*, y_*)$ at the site where the wave 
pattern is formed, the amplification of noise and the nature of formed wave are 
insensitive to the input value. Hence the nature of generated wave depends little on the 
input. 
\end{description}

\begin{figure}[hbtp]
\begin{center}
(a)\includegraphics[scale=0.28]{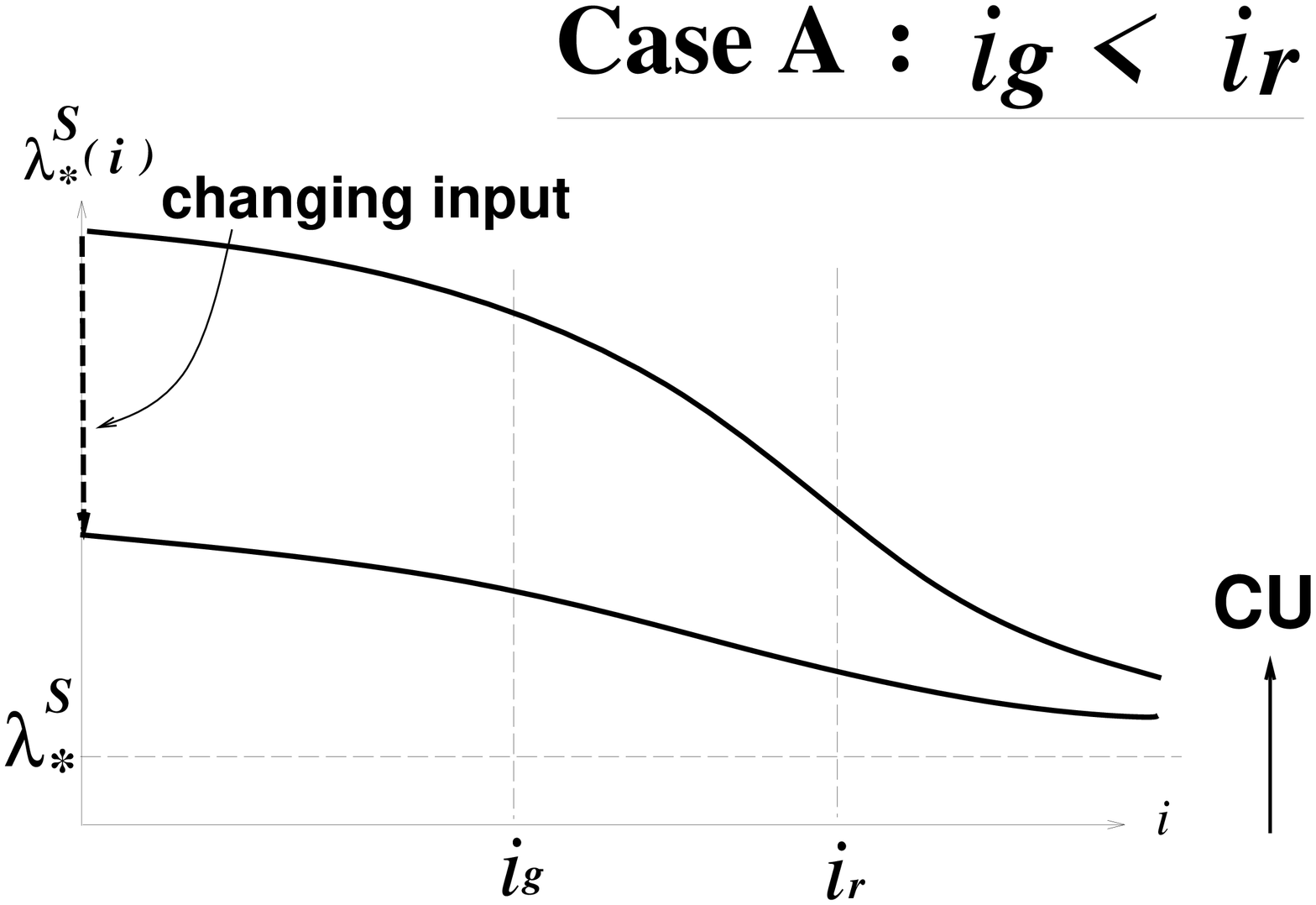}
(b)\includegraphics[scale=0.28]{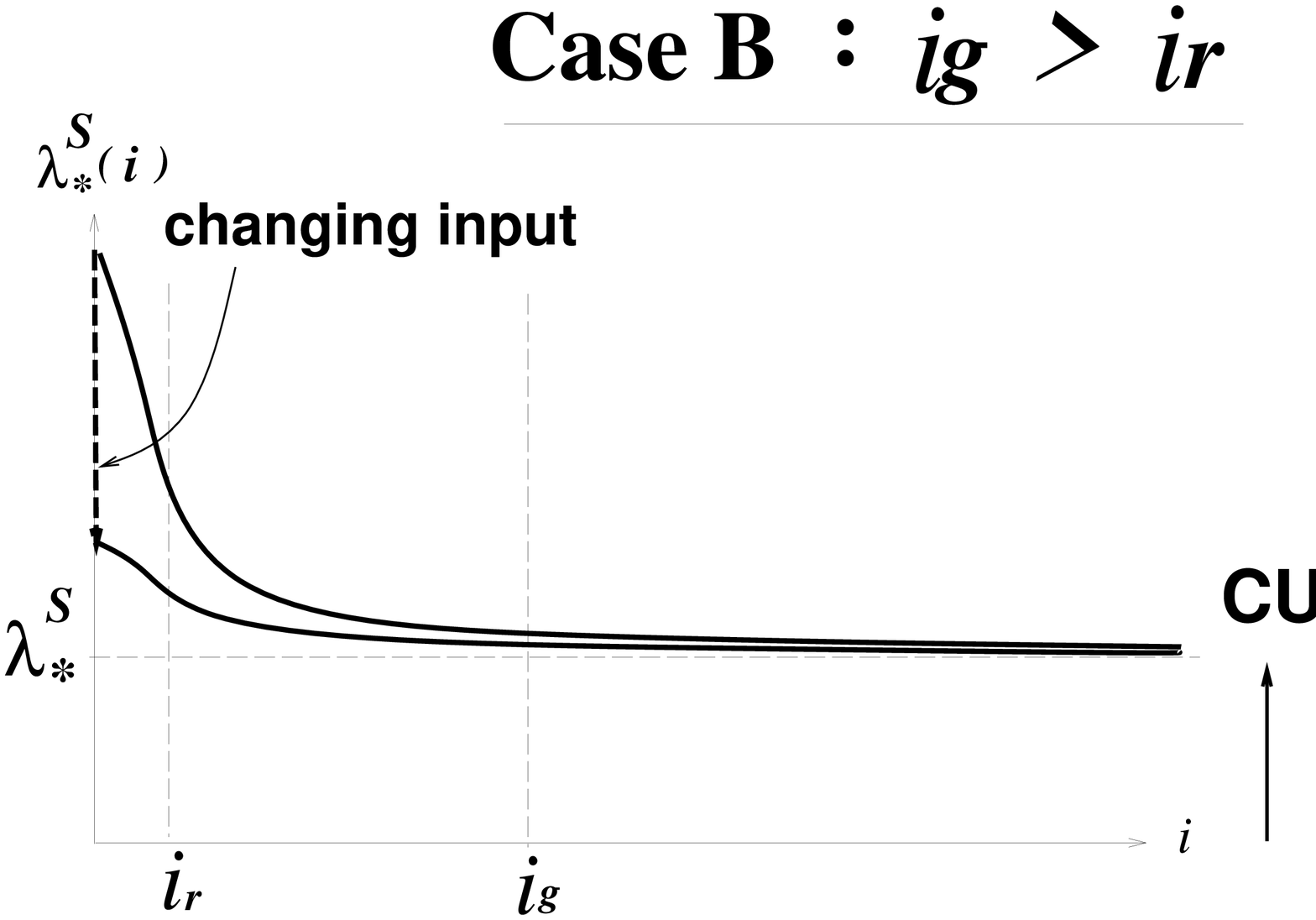}
\caption{Schematic diagram of the analogue change. Cases with a clear analogue 
change (a), and without it (b). }
\label{fig:AnalogueChange}
\end{center}
\end{figure}

To write down the condition for the analogue dependence for the Type2, we need to add 
the condition for the formation of wave (discussed in \S\ref{Mechanism of digital 
change}), while for the Type3 the wave is always formed. Hence the conditions of the 
analogue dependence on the input are summarized as

\begin{equation}
\left\{\begin{array}{ll}
\mbox{Type2 \ : \ }(i_g < i_r )\cap (i_g \stackrel{<}{\sim} i_u ) \\
\mbox{Type3 \ : \ }i_g < i_r \
\end{array}\right. 
\end{equation} 
In Fig.\ref{fig:AnalogueChange} the above mechanism of the analogue change is 
shown schematically.

Note again that our mechanism for input dependence works universally in one-way 
coupled dynamical systems, irrespective of the choice of our specific model. In our model, 
the analogue change is represented as the frequency of oscillations. This transformation 
of the input to the frequency can be model specific. In general, there may be a variety of 
ways of transformation of input values to dynamics at the down-flow. Still the argument 
presented here is applicable to each case.

\section{Noise effect to input dependence}
In the previous section, the mechanism of input dependence is explained from the 
spatial change of $\lambda^S(i)$. In this section, mechanism of input dependence is 
discussed from a different angle, that is the relation between convective instability and 
noise (fluctuation) intensity. It will be shown that the input dependent dynamics at the 
down-flow is seen only within some range of noise intensity. 

\begin{figure}[hbtp]
\begin{center}
\includegraphics{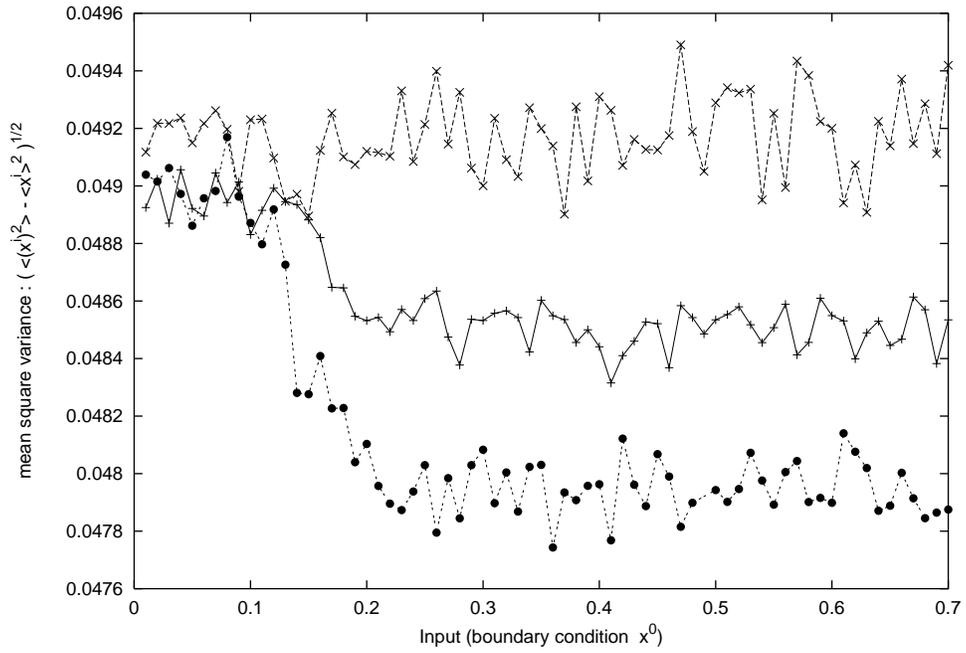}
\caption{Mean square variance $( <(x^i)^2>-<x^i>^2 )^{1/2}$ is plotted against the 
change of input values, for 3 different values of noise amplitudes. $+$ 
($\sigma = 1.0 \times 10^{-7}$) $\rightarrow$$\bullet$ ($\sigma = 3.0 \times 10^{-3}$) 
$\rightarrow$$\times$ ($\sigma = 8.0 \times 10^{-2}$). The parameters corresponds 
to Type3, $a = 2.7, b = 10.7, c = 4.8, d = 9.4, K = 0.015, \epsilon = 4.5$. }
\label{fig:xf-bunsan-analogue}
\end{center}
\end{figure}

Fig.\ref{fig:xf-bunsan-analogue} shows our numerical results on the input dependence 
of the variance of $x^i(t)$ at the down-flow for different amplitudes of noise, where the 
parameters are fixed so that the system belongs to Type3. It is seen that the mean 
square variance has input dependence only in medium intensity ($\bullet$) of the 
noise. For too large ($\times$) intensity of the noise, the variance does not show input 
dependence, while the dependence gets weaker as noise amplitude is decreased ($+$). 
``Suitable'' range of noise intensity is required to have analogue change on the input. 

\subsection{disappearance of input dependence at a low noise regime}

The mechanism of the disappearance of input dependence at a low noise regime is 
straightforwardly explained with the argument of the previous section. According to 
eq.(\ref{eq:i_g}), $i_g$ gets larger with the decrease of the noise, and the relationship 
between $i_g$ and $i_u$ or $i_r$ changes. With the decrease of $\sigma$, the following 
two changes are possible. 

\begin{enumerate}
\item Disappearance of digital change (for Type2):

With the decrease of $\sigma$ by fixing the input, $i_g$ can be larger than $i_u$, and 
the transition from the periodic to stochastic phase, and then to fixed point phase occurs. 
The periodic phase no longer appears irrespective of the input values (within their 
allowed range in our model) for small $\sigma$. Hence digital dependence on inputs 
disappears. 

Fig.\ref{fig:IUIDbunsan}(a) shows input dependence of the variance at the down-flow 
dynamics with the change of the noise amplitude, while in Fig.\ref{fig:IUIDbunsan}(b) 
input dependence of $i_g$ and $i_u$ is plotted. It is demonstrated that the digital 
change requires larger input values as the noise amplitude is decreased. 

\begin{figure}[hbtp]
\begin{center}
(a)\includegraphics[scale=0.49]{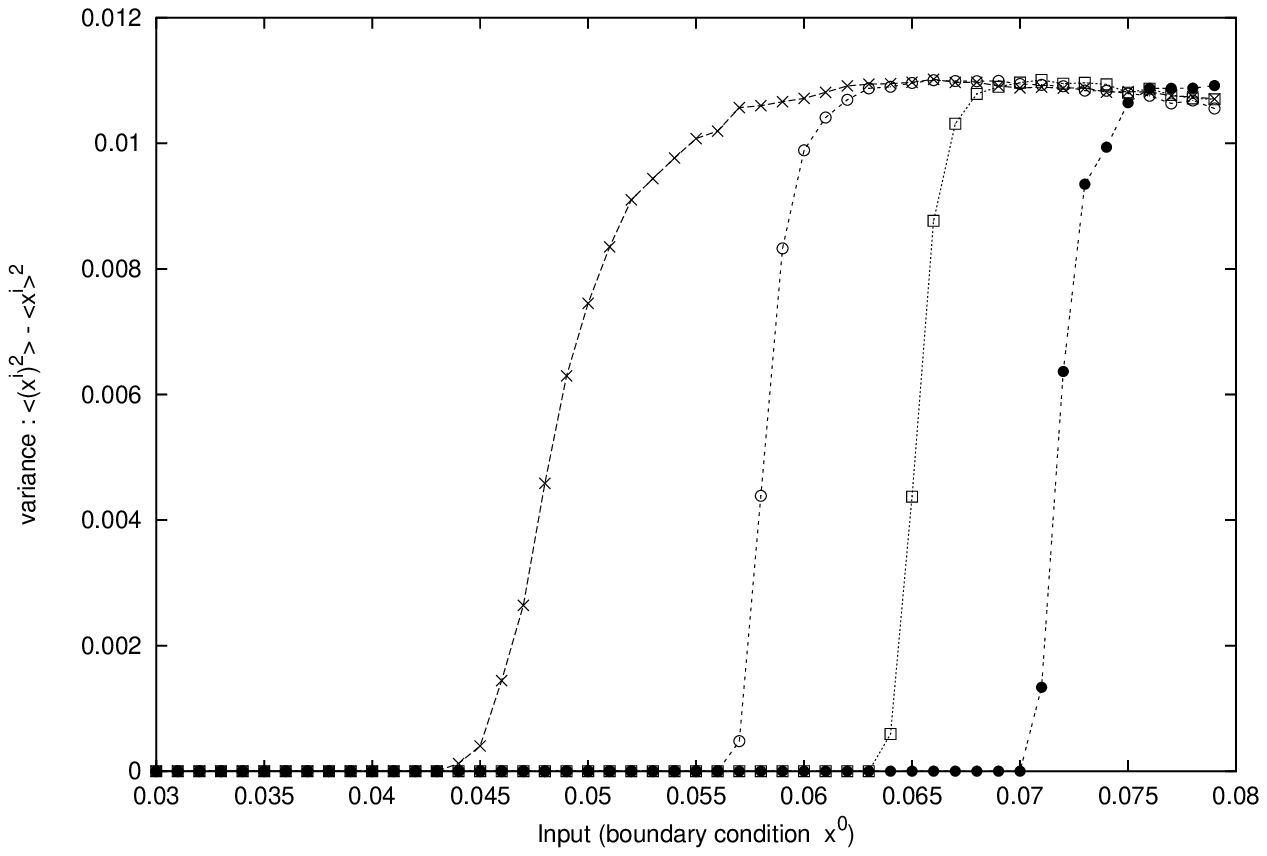}
(b)\includegraphics[scale=0.49]{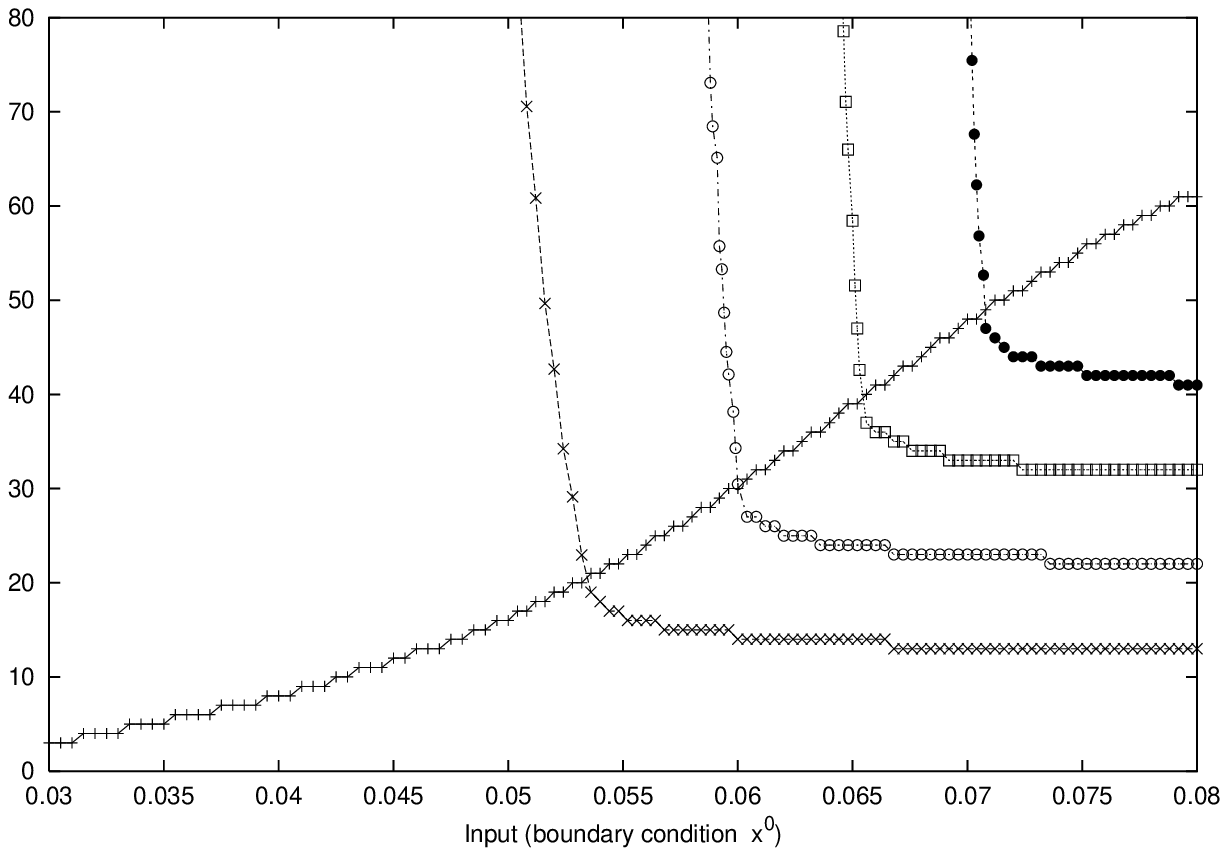}
\caption{Input dependence of the variance at $i=80$ (a)and of $i_u$ ($+$) and $i_g$ (b) are 
plotted. The noise amplitudes are $\sigma$ = $10^{-3} (\times)$, $10^{-5} (\circ)$, 
$10^{-7} (\Box)$, $10^{-9} (\bullet)$ for each figure (Note $i_u$ does not depend on 
$\sigma$). The smaller the noise amplitude is, the harder the periodic phase appears 
within allowed range of inputs. Parameters $a, b, c, d, K, \epsilon$ are same as in Fig.4. }
\label{fig:IUIDbunsan}
\end{center}
\end{figure}

\item Disappearance of analogue change (for Type2 or Type3):

With the decrease of $\sigma$, $i_g$ increases until it gets larger than $i_r$, and the 
change from CaseA to CaseB in \S\ref{Mechanism of analogue change} follows. Thus 
analogue dependence on the input fades out. 

\end{enumerate}

The lower limit for the noise $\sigma_S$ to have the input dependence is 
straightforwardly estimated from the above argument. The noise intensity must satisfy
\begin{equation}
i_g <\left\{
\begin{array}{ll}
i_r \mbox{ \ (analogue change)}\\i_u \mbox{ \ (digital change)}\
\end{array}\right. 
\end{equation}
By introducing the average spatial instability $\overline{\lambda^S_*} = 
(1/i_g)\sum_i^{i_g}\lambda^S_*(i)$, and using the expression eq.(\ref{eq:i_g}) for 
$i_g$, we obtain

\begin{equation}
\sigma > \sigma_S =
\left\{
\begin{array}{ll}
exp(-\overline{\lambda^S_*}i_r) \mbox{ \ (analogue change)}\\
exp(-\overline{\lambda^S_*}i_u) \mbox{ \ (digital change)}\
\end{array}\right. 
\end{equation}

\subsection{collapse of input dependence for a strong noise intensity}

The collapse of input-dependence at strong noise is due to a different mechanism. When 
the noise intensity is too large, the wave, once formed, can be destroyed due to the noise 
at the down-flow\footnote{The noise here plays the same role as that added to 
information channel in information theory. }. The generated oscillation is distorted by 
noise and the transmission to down-flow does not work well. In particular, for the Type2 
case, due to the convective stability of the fixed point at the down-flow, the dynamics 
can stay near the fixed point, after the collapse of the limit cycle. Thus, the input 
dependence between the periodic and fixed point states fades out. There is an upper 
limit of noise intensity $\sigma_L$ such that the input dependence disappears 
for $\sigma > \sigma_L$. 

This upper limit $\sigma_L$ cannot be estimated by the convective stability of the 
fixed point. It is a nonlinear effect of noise around the limit cycle oscillation, for which 
no theory is available as yet.

\section{Discussion and conclusions}

In the present paper we have reported boundary (input) dependence in a one-way 
coupled differential equation, and presented a general mechanism for it. It is shown 
that the spatially dependent convective instability and noise lead to such boundary 
dependence. 

Boundary condition dependence has also been studied in dissipative 
structure\cite{Prigogine}. However our boundary condition dependence is clearly 
distinguished from earlier studies, and is essential in nature. 

The most important difference is dependence on the system size. The boundary 
condition dependence previously studied is nothing but a finite size effect. The 
dependence disappears as the size gets larger. On the other hand, our boundary 
condition dependence is not due to such a finite size effect, but remains even in the 
infinite size limit. Difference of dynamics, created by the boundary (input) difference, is 
fixed to the down-flow element $i$ and is kept even for the limit $i \rightarrow \infty$. 
Also it should be noted that our boundary dependence is independent of initial 
conditions, and is stable against perturbations. 

Furthermore our mechanism supporting the boundary dependence is novel, and should 
be distinguished from that by the previous studies. In our case, it is originated in the 
spatial amplification of fluctuation by the convective instability, in contrast with the use 
of a stable system in the earlier studies \cite{Prigogine}. The amplification rate of 
fluctuation, characterized by the (local) convective Lyapunov exponent $\lambda_v(i)$, 
depends on space, and accordingly on the boundary (input). After the amplification, the 
dynamics is stabilized at the down-flow. Once stabilized, the dynamics stays as an 
attractor. Dynamics before the achievement of the spatial convergence can depend on 
the boundary. On the other hand, the generated difference at the down-flow dynamics 
cannot be changed any more and is fixed, due to the stable dynamics there. 
Hence the boundary dependence is fixed at down-flow. 

We have shown that there are two types of boundary (input) dependence. 
One is the threshold-type (bifurcation-like) change, 
leading to qualitative difference in the down-flow dynamics. 
In our model, bifurcation from the fixed point, to the stable stochastic 
oscillation, and then to the stable periodic oscillation is found with the change of the 
boundary value. The other is continuous (analogue) change of down-flow dynamics, 
depending on the boundary value. In our example, the frequency of pulses can 
continuously change with the boundary value. 

These two types of boundary dependence are quantitatively analyzed by introducing 
three characteristic length scales based on the local convective Lyapunov exponent 
$\lambda_v(i)$. The first one is the length scale $i_g$ necessary for the fluctuation to 
be amplified to the macroscopic order. The second is the length scale $i_u$ where the 
convective instability of the fixed point disappears and the dynamics becomes 
convectively stable. The third is the length scale $i_r$ characterizing the spatial 
convergence of $\lambda_v(i)$. 

For the bifurcation-like change, it is crucial if the amplification of fluctuation is 
completed before the fixed point dynamics becomes stable. Otherwise, the down-flow 
dynamics remains at the fixed point motion. Thus the condition for the qualitative 
change against the boundary is determined by $i_g \stackrel{>}{<} i_u$. On the other 
hand, the analogue change depends on if the amplification is completed while the 
spatial dependence of the convective stability is significant, and thus is judged by the 
condition $i_g \stackrel{>}{<} i_r$. Note that these conditions also imply that such 
boundary dependence is seen only within some range of noise amplitude $\sigma$, 
because $i_g \propto log \sigma$ from eq.(\ref{eq:i_g}). 

Since our mechanism is expressed in universal terms in dynamical systems, the input 
(boundary) dependence is expected to be observed generally in a spatially extended 
dynamical system, as long as the spatially inhomogeneous convective instability exists. 
It is interesting to search for such boundary dependence for open-fluid flow, optical 
system \cite{Ikeda-Otsuka}, coupled map lattice\cite{KK0, Frederick-KK}, and so forth, 
and to check if the relationship among $i_g$ and $i_u$ and $i_r$ are satisfied. 

Our motivation for this study is originated in biological signaling problems. Then what 
implications are drawn from our result to such problems? 

In our study the input information is represented by the boundary value, while the 
response is given by a state of the down-flow dynamics. At the region $i 
\stackrel{<}{\approx} i_g$ required for the wave formation, the input difference is 
amplified by the convective instability. 
The information on the dynamics at this region is 
translated to the response, through the site dependent $\lambda_v(i)$. If 
the conditions among $i_g$, $i_u$, and $i_r$ are satisfied, input information can be 
translated to output information through the chemical dynamics at the signaling 
pathway. 

Now let us come back to the questions raised in \S 1. The first question is on the reason 
for the length and complication of the signaling pathway, to which our results have 
some implication. According to our mechanism, we need several sites (larger than 
$i_g$ and $i_u$) to have input dependence. This number of sites depends on the 
parameter, but cannot be too small, as long as the dynamics is not too convectively 
unstable. Hence the pathway must have a sufficient length, to have input dependence. 
Of course the signal pathway does not consist of a single chain of reaction, but several 
chains are mutually influenced. Although extension of our study to a multiple chain 
case remains as a future problem, it is expected that such complication can afford an 
effectively long chain required for the input dependence. 

The second question on the suitable response to inputs is answered by the above 
translation of the input (boundary) to response (down-flow dynamics) through spatially 
dependent convective instability $\lambda_v(i)$. In particular, the digital (bifurcation-
like) dependence provides a response with some input threshold, while the analogue 
dependence provides the translation from input concentration to response frequency. 
Both the dependence are essential to neural and sensory responses. It is also interesting 
to note that our signal transmission mechanism is different from that given by 
Hodgkin-Huxley equation for neural signal transmission\cite{Hodgkin-Huxley}. 

The third question on the robust signaling mechanism under thermodynamic 
fluctuation is most clearly answered by our mechanism. Indeed our input dependence 
works in the presence of noise, or rather the noise amplitude $\sigma$ has to satisfy 
$\sigma_L > \sigma > \sigma_S$. Note that the noise is inevitable in a cell system by 
thermal fluctuation and also due to a relatively small number of signaling molecules. Indeed the 
number of most signal molecules is around $N \sim 1000$. Thus the noise of the order 
$1/\sqrt{N} \sim 0.03$ should exist. The condition $\sigma_L > \sigma > 
\sigma_S$ for the input dependence may be suggestive in thinking why most cell 
response works only within some temperature range, and works in a system of small 
number of (signaling) molecules. 

Of course several studies have to be pursued in future, not only as dynamical systems 
but also for the application of our viewpoint to biological signaling phenomena. The 
following list is under  current investigation. 

\begin{enumerate}
\item Extension to a case with complex spatial dynamics of $\lambda_v(i)$: In our 
model $\lambda_v(i)$ monotonically relaxes to $\lambda_v(\infty)$. 
In general, the relaxation can be oscillatory or have chaotic transient, as is found in a 
one-way coupled map lattice (OCML)\cite{Fuji-KK}. In such case, complex input 
dependence may be expected. 

\item Extension to a case with a bi-directional or symmetric coupling: Although the 
present uni-directional coupling case provides a most straightforward example to see 
the convective instability, such instability also exists in a bi-directional coupling case, 
and is characterized by the co-moving Lyapunov exponent $\lambda_v(i)$ \cite{KK-Co-move}. 
Indeed some preliminary studies show that the boundary condition 
dependence of our type also exists in such case. 

\item Extension to a system with several CS attractors at $i \rightarrow \infty$. In 
this case choice of an attractor can depend on inputs. Complicated dependence on input 
values may exist in a similar manner with fractal or riddled basin structure for a 
multiple attractor system. 

\item Extension to a non-constant input (boundary). If the boundary oscillates in time, 
for example, this temporal information of the input (boundary) can be translated to 
response dynamics. Indeed, in a OCML\cite{Frederick-KK}, the input oscillation is 
selectively transmitted to down-flow depending on its period. 

\item Extension to a multiple chain system with multiple inputs. In a signaling process 
in a cell, several pathways exist, which interact with each other. With multiple 
boundary values (inputs), a new type of boundary dependence can be expected, such as 
a response depending on combination of several inputs.

\end{enumerate}

{\sl acknowledgments}

The authors are grateful to T.Yomo, T.Shibata, and I.Tsuda for stimulating discussions. 
The work is partially supported by Grant-in-Aids for Scientific Research from the 
Ministry of Education, Science, and Culture of Japan.

\appendix

\section{calculation of local co-moving Lyapunov exponent of the fixed point $(x^i_*, 
y^i_*)$}

Since the fixed points change their value by sites in our model, the conventional 
method\cite{KK-Crutch} to calculate co-moving Lyapunov exponent has to be extended. 
To measure the amplification rate of perturbation at one site $i$, we consider a system 
where $(x^j, y^j) = (x^i_*, y^i_*)$ for $j \geq i$ and compute the amplification rate per 
lattice point. Let us consider the evolution equation,

\begin{equation}
\dot{\vec{z}}^j(t) = \vec{h}^j(\vec{z}^j(t), \vec{z}^{j-1}(t))
\label{eq:appendix-model}
\end{equation}
with $\vec{h}^j(\vec{z}^j, \vec{z}^{j-1}) = (f(\vec{z}^j, \vec{z}^{j-1}), g(\vec{z}^j, 
\vec{z}^{j-1}))$

The displacement follows the equation, 

\begin{equation}
\delta \dot{\vec{z}}^j(t) = 
\frac{ \partial{\vec{h}}^j(\vec{z}^j, \vec{z}^{j-1}) }{ \partial{\vec{z}^j }} \delta 
\vec{z}^j(t)
+ \frac{ \partial{\vec{h}^j(\vec{z}^j, \vec{z}^{j-1})} }{ \partial{\vec{z}^{j-1} }} \delta 
\vec{z}^{j-1}(t)
\end{equation}

We solve the equation for $\vec{z}^j = (x^i_*, y^i_*)$ for $j \geq i$ and $\forall t$ with 
the initial condition $\delta \vec{z}^j(0) = \delta_{0, j} \delta$. The solution can be 
written as 

\begin{equation}
\delta \vec{z}^j(t) \equiv J^j(t) \delta \vec{z}^0(0)
\label{eq:appendix-define-matrix}
\end{equation}
where
\begin{equation}
\dot{J}^j(t) = \frac{ \partial{\vec{h}}^j(\vec{z}^j, \vec{z}^{j-1}) }{ \partial{\vec{z}^j }} 
J^j(t)
+ \frac{ \partial{\vec{h}^j(\vec{z}^j, \vec{z}^{j-1})} }{ \partial{\vec{z}^{j-1} }} J^{j-1}(t)
\label{eq:appendix-delZ}
\end{equation}

Here, ${J}^j(t)$ is a $2 \times 2$ matrix, and is obtained by  solving  
eq.(\ref{eq:appendix-delZ}) under the initial condition $\delta\vec{z}^j(0) = \delta_{0,j}\delta$. 
By defining $e^{\Lambda_1^{j}(t)}$ and $e^{\Lambda_2^{j}(t)}$ as absolute values of 
the eigen values of $J^{j}(t)$ ($\Lambda_1^{j}(t) \geq \Lambda_2^{j}(t)$) we obtain, 
\begin{equation}
\lambda_v(i) \equiv \lim_{t \rightarrow \infty} \frac{1}{t} \left. 
\Lambda_1^{j}(t)\right|_{j = vt}
\end{equation}

\end{document}